\documentclass[%
 preprint,
 amsmath,amssymb,
 aps,
 prb,
 longbibliography,
 lengthcheck,%
]{revtex4-1}

\usepackage{graphicx}
\usepackage{bm}
\usepackage{mathtools}
\usepackage[mathlines]{lineno}

\newcommand{\Q}{\mathbf Q}
\newcommand{\Sb}{\mathbf S}

\newcommand{\rr}{\mathbf r}

\DeclareMathAlphabet{\mathcalligra}{T1}{calligra}{m}{n}

\usepackage{floatrow}
\newfloatcommand{capbtabbox}{table}[][\FBwidth]

\usepackage{blindtext}

\usepackage{hyperref}

\begin{document}


\title{Spin excitation spectrum of high-temperature cuprate superconductors from finite cluster simulations}

\author{Oleg Lychkovskiy$^{1}$}

\author{Boris V. Fine$^{1,2}$}
\affiliation{
$^1$
Skolkovo Institute of Science and Technology,
Nobel street 3, Moscow  143026, Russia}
\affiliation{
$^3$
Institute for Theoretical Physics, University of Heidelberg,
Philosophenweg 12, 69120 Heidelberg, Germany}

\date{\today}


\begin{abstract}
A  cluster of spins $1/2$ of a finite size can be regarded as a basic building block of a spin texture in high-temperature cuprate superconductors. If this texture has the character of a network of weakly coupled spin clusters, then spin excitation spectra of finite clusters are expected to capture the principal features of the experimental spin response. We calculate spin excitation spectra of several clusters of spins $1/2$ coupled by Heisenberg interaction. We find that the calculated spectra exhibit a high degree of variability representative of the actual phenomenology of cuprates, while, at the same time, reproducing a number of important features of the experimentally measured spin response. Among such features are the spin gap, the broad peak around $\hbar \omega\simeq (40 - 70)$ meV and the sharp peak at zero frequency.
The latter feature emerges due to transitions inside the ground-state multiplet of the so-called ``uncompensated'' clusters with an odd number of spins.
\end{abstract}

\maketitle

\section{Introduction}


Neutron scattering experiments in high-temperature cuprate superconductors reveal  intricate patterns of magnetic spin response.\cite{fujita2011progress,brooks2007handbook,supplement} Recently some of these findings were corroborated by the resonant inelastic X-ray scattering (RIXS) experiments.\cite{miao2017high}
A broad range of theoretical aproaches of varying degree of sophistication, including those based on Hubbard,\cite{seibold2005magnetic,seibold2006doping,kung2015doping} $t-J$,\cite{onufrieva2002spin,sushkov2005theory,james2012magnetic,eremin2012dual} Heisenberg,\cite{uhrig2005magnetic,vojta2004magnetic,yao2006magnetic,mang2004spin,andrade2012disorder} spin-fermion\cite{abanov1999relation}  models  and their extensions have been employed in order to reproduce this spin response.
At present, however, there is no consensus  about the starting set of assumptions to describe the available phenomenology in  cuprates. Here one faces dilemmas between the pictures of itinerant and localized spins, between including inhomogeneous spin textures at the level of model assumptions or obtaining these textures dynamically from the spin susceptibility of the homogeneous parent state. When inhomogeneous textures are assumed, one has a choice of either stripe or checkerboard patterns, or more disordered  "Swiss-cheese"-type of textures.\cite{bianconi2013quantum,campi2015inhomogeneity} In the latter case, one expects that the antiferromagnetic order is retained locally  within finite spin clusters (domains, puddles etc), at least approximately, while intercluster correlations fade away with doping.  In general, irrespective of the initial set of assumptions, one can reasonably expect that the spin response at sufficiently high frequencies for infinite systems and for finite parts of these systems would be approximately the same. If, however, the spin texture has the character of a network of weakly coupled spin clusters, the cluster calculations can also capture important experimentally observed features at low frequencies such as the onset of a spin gap. Such a texture can, indeed, emerge as a possible realization of Coulomb-frustrated phase separation.\cite{emery1993frustrated,fine2008phase}
In this work, we investigate spin responses of finite clusters of spins $1/2$ described by the Heisenberg model, in an attempt to test whether the experimental phenomenology can be understood from a unified perspective of a finite-cluster simulations. We find that such an approach is indeed quite promising.

We start in the next section with a brief overview of the main features of spin response observed in cuprates along with related theoretical concepts and considerations. In Sec. \ref{sec: model}, we formulate our cluster model and discuss its general properties. Several clusters are considered, with sizes and shapes motivated by theoretical considerations and experimental data. In Sec. \ref{sec: results}, we present the results of numerical simulations  and discuss their relation to the experimental data.  The simulated magnetic response from these clusters exhibits several features observed experimentally across a wide range of cuprate compounds.
Discussion and conclusions are finally presented in Sec. \ref{sec: conclusions}.

\section{Main features of  spin response in cuprates \label{sec: experiments}}

\begin{figure*}[t]
\includegraphics[width = \textwidth]{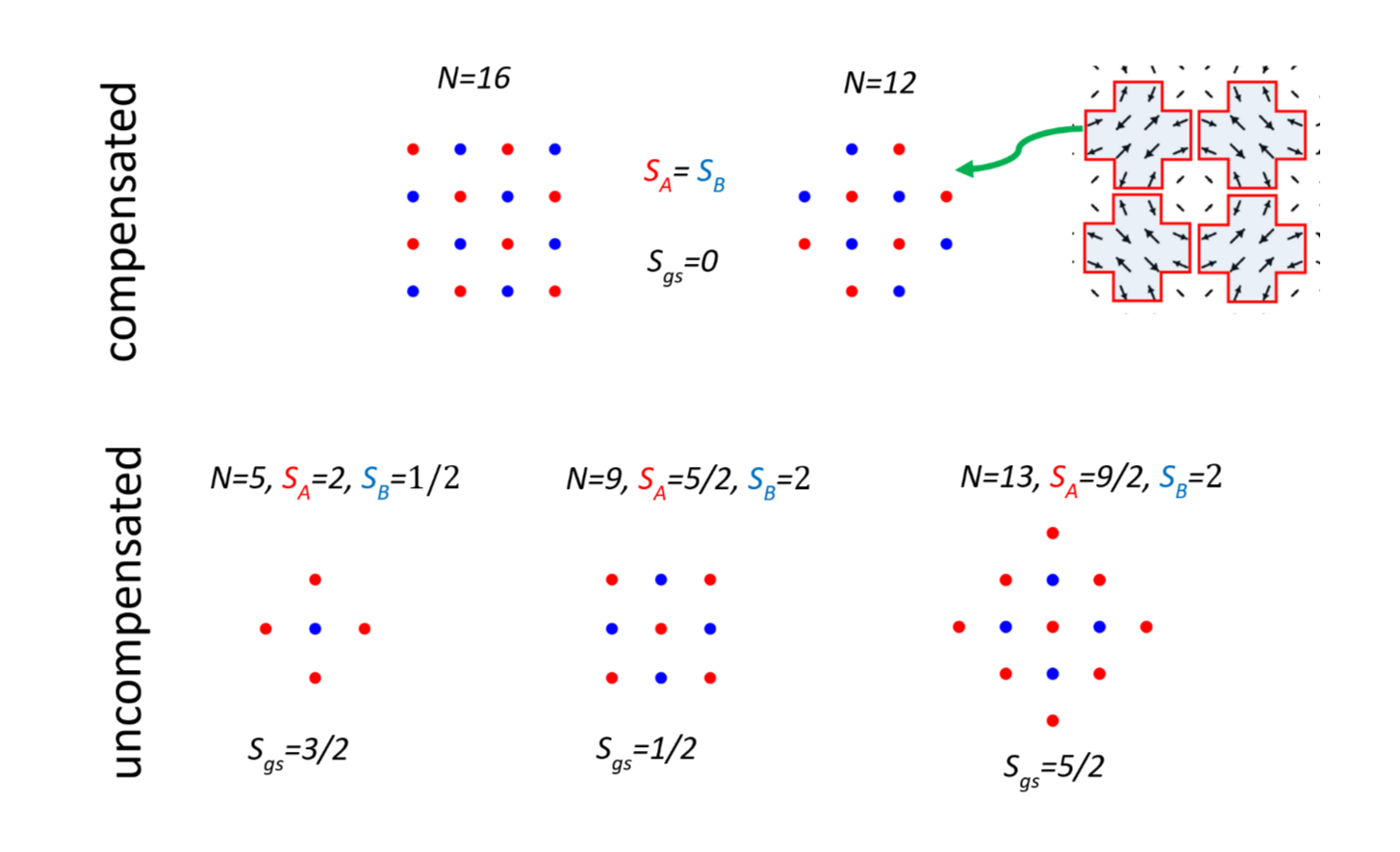}
 \caption{
Compensated (upper row) and uncompensated (lower row) clusters considered in the paper with the total number of spins $N=5,9,12,13,16$. Sublattices A and B of each cluster are shown in red and blue, respectively.
$S_{A}$, $S_{B}$ and $S_{gs}$ are, respectively, the maximal total spins of the sublattices $A$ and $B$ and the total spin of the ground state of the whole cluster.
Spin-vortex texture shown next to the cluster with $N=12$  was considered in Refs. \onlinecite{fine2007magnetic,dolgirev2017pseudogap,bhartiya2017superconductivity} --- it includes the cluster with $N=12$ as the basic building block.
 \label{fig:clusters}
 }
\end{figure*}

We consider the spin excitation spectrum probed by neutron scattering experiments. This technique allows one to measure the imaginary part of the dynamical susceptibility $\chi''(\Q,\omega)$. In the low-temperature limit, it is defined as\cite{brooks2007handbook}
\begin{align}\label{S(Q,w)}
\chi''(\Q,\omega) \equiv & \sum_{\alpha,\beta}\left(\delta_{\alpha\beta} - Q_\alpha Q_\beta/\Q^2 \right) \nonumber\\
& \times \frac1{2\pi} \int dt e^{-i \omega t} \sum_{\rr,\rr'} e^{i \Q(\rr-\rr')} \langle S_{\rr'}^\alpha (0) S_{\rr}^\beta (t) \rangle,
\end{align}
where $\Sb_{\rr}$ is the operator of the spin $1/2$ in the lattice site with the coordinate $\rr$ and the quantum mechanical average, $\langle S_{\rr'}^\alpha S_{\rr}^\beta \rangle$, is taken over the subspace of ground states.  We put $\hbar=1$.
One can also define the $\Q$-integrated local susceptibility $\chi''(\omega)$as
 \begin{equation}\label{S(w)}
\chi''(\omega)\equiv \int d\Q\, \chi''(\Q,\omega).
\end{equation}


A large body of experimental work has been done on the neutron scattering in high-temperature cuprate superconductors  (see e.g. Refs. [\onlinecite{fujita2011progress}],[\onlinecite{brooks2007handbook}] for reviews). 
Detailed experimental data are available for a number of compounds, for which sufficiently large single-crystal samples can be produced. Below we list the principal features of these data. For convenience of the reader we have collected in the Supplemental Material experimental plots which demonstrate these features. We refer to these plots, Figs. \ref{fig:experimental} and \ref{fig:hourglass}, on several occasions below.


{\it Spin gap vs a peak  at $\omega\simeq 0$.} The spin response around zero frequency comes in two variations. On the hole-doped side, underdoped lanthanum cuprates exhibit a narrow peak at zero frequency followed by a frequency range with suppressed response and then one or more somewhat broader peaks\cite{tranquada2004quantum,lipscombe2009emergence} [see Fig. \ref{fig:experimental} (a) and (e) in \onlinecite{supplement}], while optimally doped and overdoped lanthanum cuprates \cite{vignolle2007two,lipscombe2007persistence} as well as other not too underdoped  cuprates
\cite{dai2001evolution,chan2016hourglass,chan2016commensurate,stock2010effect,xu2009testing}
generally exhibit a spin gap, defined as a frequency threshold below which $\chi''(\omega)$ vanishes [see \onlinecite{supplement},  Fig. \ref{fig:experimental} (b)--(d),(f),(g)]. This spin gap varies from a few meV to tens of meV. In the case of optimal doping, it appears to correlate with the critical temperature.\cite{brooks2007handbook} In electron-doped cuprates, a peak at $\omega\simeq 0$ was also reported in ref. \onlinecite{wilson2006high}.  However this peak is possibly separated from $\omega=0$ by a small gap, see Fig. \ref{fig:experimental} (h) in \onlinecite{supplement}.  It should be noted that, even in the compounds where spin gap is apparently absent, the magnetic response at low frequencies is very different from that of parent antiferromagnets, where it is dominated by spin waves.\cite{coldea2001spinwaves,yao2006magnetic}


From the theoretical standpoint, a spin gap or its absence appears to be a rather delicate issue. While the gap is absent in a Heisenberg model for spins $1/2$ both on a square lattice and in a one-dimensional chain, it can appear in spin ladders with even number of legs.\cite{barnes1993excitation}  This supports the idea that spatial inhomogeneities can be responsible for the gap. On the other hand, the gap in spin ladders is not particularly robust and can vanish for ladders with next-nearest and four-spin interactions.\cite{wang1999exactly}


{\it Hourglass and wine glass structure of $\chi''(\Q,\omega)$.}
In doped cuprate superconductors the antiferromagnetic peak splits in a way which often, but not always, gives rise to the celebrated hourglass structure -- a feature common for a variety of compounds,\cite{fujita2011progress,brooks2007handbook} see Fig. \ref{fig:hourglass} (a) in \onlinecite{supplement}. An alternative to the hourglass is a Y-shaped structure also known as the ``wine glass'' which has been observed in a slightly underdoped HgBa$_2$CuO$_{4+\delta}$, see Fig. \ref{fig:hourglass} (b) in \onlinecite{supplement}.
We note, however, that the upper branch of the hourglass is similar to the upper branch of the wine glass and therefore constitutes a more universal feature. This upper branch also appears in spin responses of various theoretical models, such as spin ladder and Hubbard models.\cite{barnes1994susceptibility,tranquada2004quantum,seibold2005magnetic,seibold2006doping} As for the lower branch, its theoretical description requires an accurate treatment of the lower-energy physics (for example, interstripe  interactions in the stripe paradigm) and thus is more difficult to justify.\cite{uhrig2005magnetic,sushkov2005theory,eschrig2006effect,yao2006magnetic,james2012magnetic}

{\it Broad peak in $\chi''(\omega)$ at $\omega_0=(40-70)$ meV.}  This feature is rather universal for hole-doped cuprates. It is observed in all cases shown in Fig. \ref{fig:experimental},\cite{supplement} except for the one shown in Fig. \ref{fig:experimental} (c)\cite{supplement} (the overdoped La$_{2-x}$Sr$_x$CuO$_4$ \cite{lipscombe2007persistence}). When this peak exists, its frequency $\omega_0$ corresponds to the waist of the hourglass or the bottom of the wine glass. Both the origin of this peak and its implications for the high-temperature superconductivity attracted a lot of theoretical and experimental attention.\cite{eschrig2006effect}
From the theoretical point of view this peak often emerges as another facet of the spin gap.\cite{dai2001evolution,uhrig2005magnetic,seibold2005magnetic,seibold2006doping,eschrig2006effect,james2012magnetic}


{\it Sharp low-frequency  peak in $\chi''(\omega)$.} This peak has been observed in underdoped, optimally doped and overdoped  La$_{2-x}$Sr$_x$CuO$_4$\cite{vignolle2007two,lipscombe2009emergence,lipscombe2007persistence,wakimoto2007disappearance} at $\omega=(7-18)$ meV, see Fig. \ref{fig:experimental} (a)--(c).\cite{supplement} Its apparent counterpart appears in La$_{1.875}$Ba$_{0.125}$CuO$_{4}$\cite{tranquada2004quantum} at
$\omega = 41$ meV and, possibly, in YBa$_2$Cu$_3$O$_{6+\delta}$\cite{stock2010effect} at $\omega = 33$ meV, see Fig. \ref{fig:experimental} (e) and (d),\cite{supplement} respectively. It is an open question whether all these peaks share the same origin.

Concluding this review, we would like to mention  that one notable dilemma unresolved so far is the dimensionality of spin modulations in cuprates, static or dynamic, if and when they exist.
Two options are commonly discussed: a one-dimensional striped spin texture (see e.g. Ref. \onlinecite{kivelson2003how} and references therein) or a two-dimensional checkerboard texture (see e.g. Refs. \onlinecite{fine2004hypothesis,seibold2011spin} and references therein).  Attempts have been made to discriminate between these options, in particular by analyzing experimentally measured splitting of the antiferromagnetic peak in $\chi''(\Q,\omega)$. However, no consensus on this matter has emerged up to date.\cite{christensen2007nature,fine2007magnetic,comin2015broken,fine2015comment,comin2016reply}

\section{Model and preliminary considerations\label{sec: model}}
\subsection{Theoretical model}

We consider five clusters of spins $1/2$ shown in Fig. \ref{fig:clusters} with the total number of spins, $N$, equal to  $5,9,12,13$ and 16. Each cluster is a piece of a square lattice where spins are coupled by the nearest-neighbour Heisenberg interaction described by the Hamiltonian
\begin{equation}
\label{H}
{\cal H}=J\sum_{\langle \rr,\rr' \rangle} \Sb_{\rr} \Sb_{\rr'},
\end{equation}
where
$\langle \rr,\rr' \rangle$ is a pair of nearest neighbours, $J$ is the coupling constant. The value of $J$ in parent cuprates lies in the range $(100 - 150)$ meV.\cite{brooks2007handbook}  For numerical estimates we take the value $J=120$ meV.

It should be noted that, since we study isolated spin clusters, no periodic boundary conditions are imposed.
The Fourier transform of a non-periodic function is continuous rather than discrete. We calculate this Fourier transform (as defined by Eq. \eqref{S(Q,w)}) numerically with a sufficiently small momentum resolution appropriate for the resulting continuous spectrum.

The linear sizes of the clusters considered  are  (2 - 4) lattice constants. These sizes roughly correspond to typical $Q$-scale of features of the experimentally observed $\chi''(\Q,\omega)$. Similar characteristic sizes of modulated spin superstructures appear in a number of theoretical proposals involving, in particular,  stripes\cite{kivelson2003how} and checkerboards.\cite{fine2004hypothesis,fine2007magnetic}

The excitation spectrum of each cluster, if described only by the Hamiltonian \eqref{H}, consists of a finite number of discrete frequencies. This means that the magnetic susceptibilities \eqref{S(Q,w)} and \eqref{S(w)} are reduced to  sums of $\delta$-functions, $\delta(\omega-\Omega_i)$, where $\Omega_i$ is the the frequency of the transition between two discrete energy levels. In reality, various effects not included in the above simple model would broaden these sharp spectral lines. These effects include fluctuations of the shape and the size of the clusters, charge carriers hopping on and off the clusters and intercluster interactions. Therefore,  we introduce a phenomenological Lorentzian broadening with  half-width $\Gamma$ by substituting
\begin{equation}
\delta(\omega-\Omega_i) \rightarrow\frac1{\pi\Gamma} \frac{1}{1+(\omega-\Omega_i)^2/\Gamma^2}.
\end{equation}
We estimate the value of $\Gamma$ as the half-width of most fine details in the experimentally measured magnetic response. Unless explicitly stated otherwise, we choose $\Gamma=0.1 J = 12$ meV, in line with experimental data obtained for La$_{2-x}$Sr$_x$CuO$_4$\cite{vignolle2007two} and YBa$_2$Cu$_3$O$_{6+\delta}$\cite{stock2010effect}. It should be kept in mind, however, that  $\Gamma$  can substantially vary from compound to compound.

Few remarks are now in order. (i) The numerical analysis of  finite cluster was employed previously as an approximate method for accessing the spin response of infinite two-dimensional spin systems (see e.g. refs. \onlinecite{manousakis1991spin}, \onlinecite{misumi2014spin}). With such a goal in mind it was natural to choose periodic boundary conditions. The agenda of the present work is  different in the sense that we assume that real spin clusters possibly exist in a sea of itinerant electrons.  Therefore we focus on the implications of the finiteness of these clusters.  In particular, we use open boundary conditions. The difference between periodic and open boundary conditions can be dramatic for small clusters, especially for uncompensated ones (defined and discussed in the next subsection).

(ii) It is known\cite{coldea2001spinwaves} that, in order to accurately describe spin excitations in undoped parent cuprate compounds, one needs to supplement the nearest-neighbour Heisenberg coupling by next-nearest-neighbour and four-spin ring exchange couplings.
While accounting for these couplings is important for quantitative description of spin excitations,
we believe that including them in our analysis  would exceed the accuracy of our basic assumption that the spin cluster consists of localized spins. Therefore keeping only the nearest-neighbour coupling in the Hamiltonian \eqref{H} should be sufficient for the qualitative and even semi-quantitative analysis  of the spin-cluster scenario.


(iii) Finally, we comment on the absolute values of the intensities of the spin response. The experimentally measured absolute intensities are known to be significantly below the theoretical estimates based on sum rules in the Heisenberg or Hubbard models.\cite{fujita2011progress,brooks2007handbook,lorenzana2005sum} A careful analysis reduces but does not completely eliminate the tension between the theory and the experiment.\cite{lorenzana2005sum} This problem can be straightforwardly resolved in the spin-cluster scenario by
choosing the concentration  of clusters in the sea of itinerant electrons sufficiently small to satisfy the experimentally measured absolute intensities. For this reason we present the  spin susceptibilities of clusters  up to an arbitrary normalization factor.

\subsection{Lieb-Mattis theorem and the spin of the ground state\label{sec: gs spin}}


The Lieb-Mattis theorem\cite{lieb1962ordering} constrains the total spin $S_{gs}$ of the ground state of a spin cluster.
It applies to bipartite spin lattices with the nearest-neighbour Heisenberg interaction \eqref{H}, i.e. lattices which can be divided into two sublattices $A$ and $B$, such that spins in each pair of nearest neighbours belong to different sublattices.  The theorem states that
\begin{equation}
S_{gs} \leq |S_A-S_B|,
\end{equation}
where $S_A$ and $S_B$ are the maximal total spins of sublattices $A$ and $B$, respectively. These maximal spins can be expressed as $S_A=\frac12 N_A$ and $S_B=\frac12 N_B$, where $N_A$ and $N_B$ are the numbers of lattice sites in the respective sublattices.



We divide all clusters considered into two categories: compensated clusters with $N_A=N_B$, and uncompensated ones, with $N_A\neq N_B$, see Fig. \ref{fig:clusters}. The total spin of the ground state of a compensated cluster is always zero due to the Lieb-Mattis theorem. In contrast, the total spin of the ground state of an uncompensated cluster can be nonzero.



It should be noted that, for finite clusters, the Lieb-Mattis theorem is also applicable in the presence of not too large next-nearest-neighbour and four-spin ring exchange couplings ({\it cf.} refs. \onlinecite{lieb1962ordering} \onlinecite{hakobyan2008antiferromagnetic}). For this reason the hierarchy of energy levels with different total spin and the structure of transitions (see below) dictated by the theorem is robust.

Another remark concerns boundary conditions. If we were to use the periodic boundary conditions, the compactification  of the  cluster with an odd number of spins would destroy the division of a lattice into sublattices and thus makes the Lieb-Mattis theorem inapplicable.

\begin{figure*}[t]
\includegraphics[width = \textwidth]{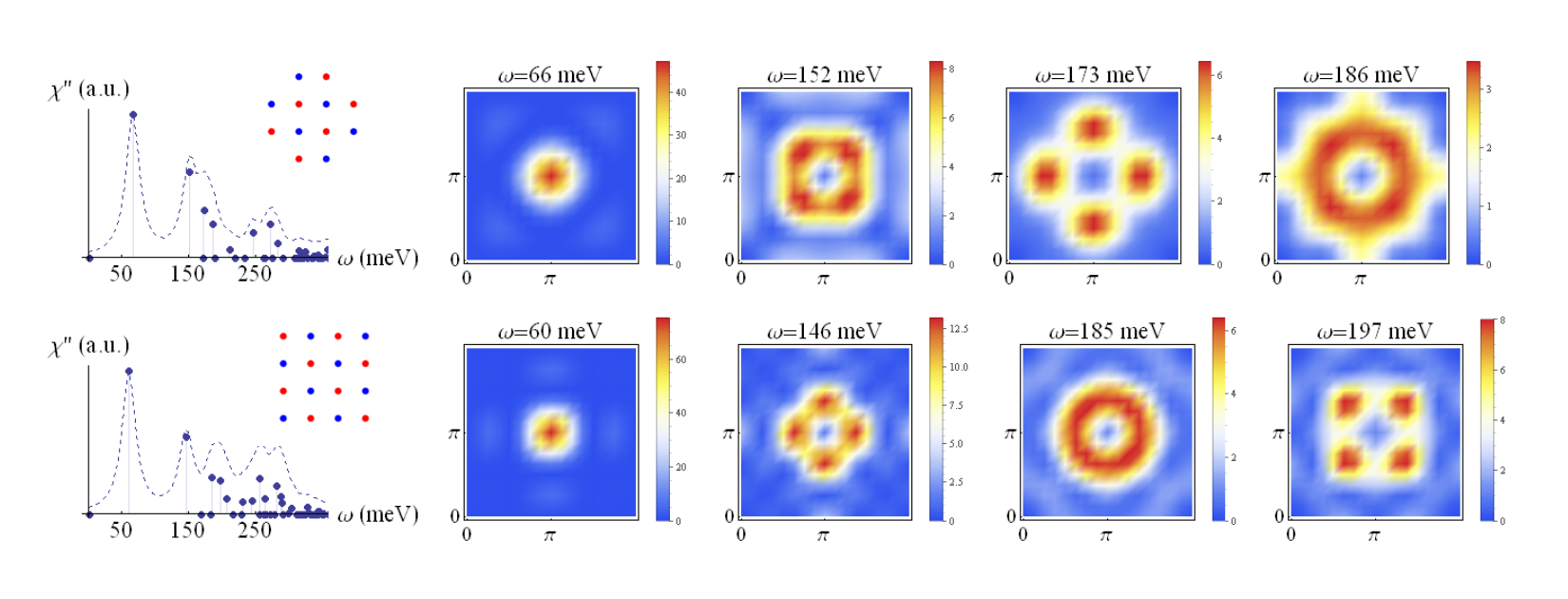}
 \caption{
Numerically calculated spin response from compensated clusters with $N=12$ (upper panel) and $N=16$ (lower panel). Plots on the left -- local susceptibility $\chi''(\omega)$;  corresponding clusters are shown in the insets. On the right are shown sections of the momentum-dependent susceptibility $\chi''(\Q,\omega)$ at fixed $\omega$.
 \label{fig:compensated}
 }
\end{figure*}

\begin{figure*}
\includegraphics[width = \textwidth]{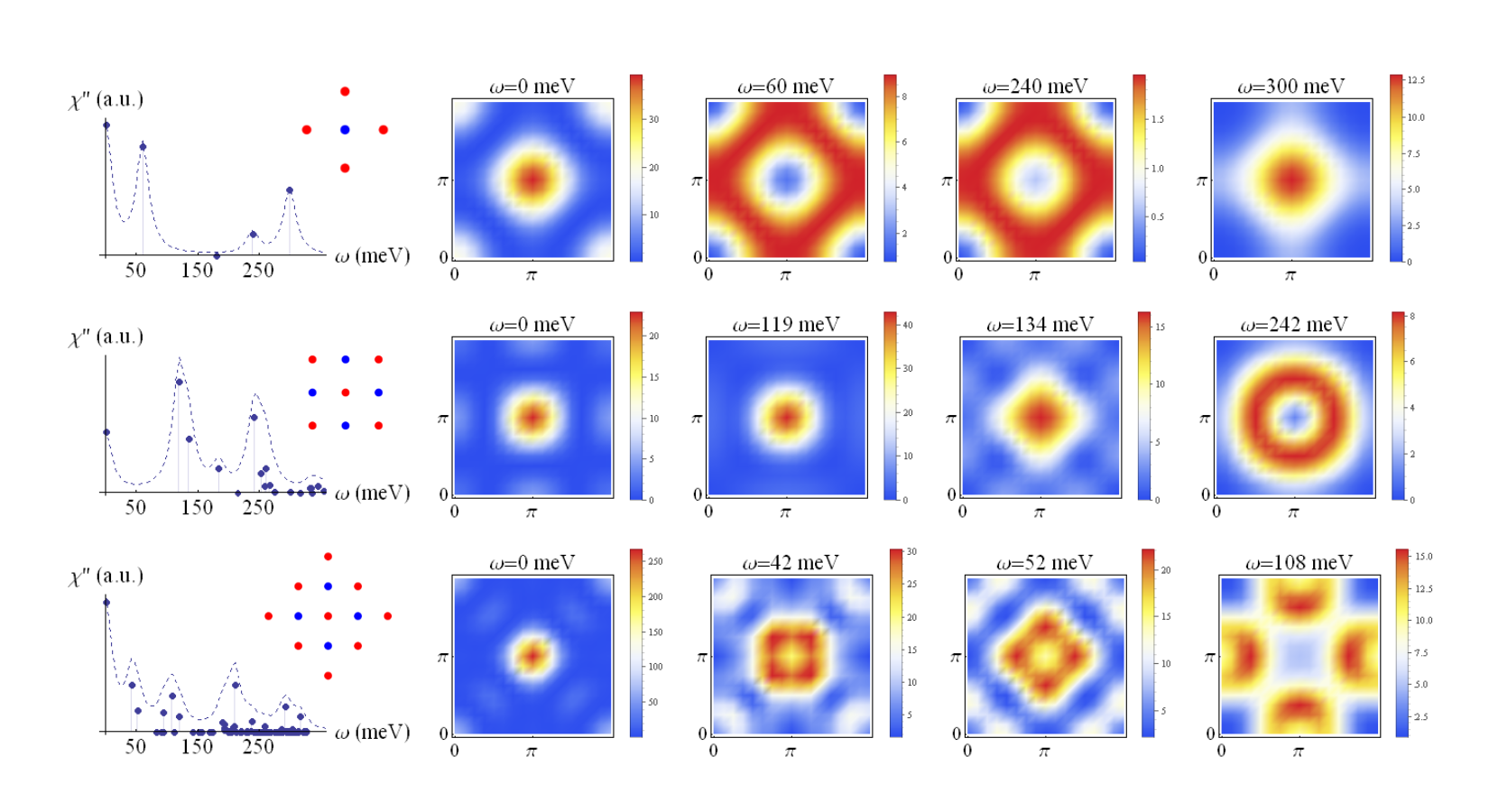}
 \caption{
Numerically calculated spin response from compensated clusters with $N=5$ (upper row), $N=9$ (middle row) and $N=13$ (lower row). Plots on the left: local susceptibility $\chi''(\omega)$ for clusters shown in the corresponding insets. Plots on the right: sections of the momentum-dependent susceptibility $\chi''(\Q,\omega)$ at fixed $\omega$.
 \label{fig:uncompensated}
 }
\end{figure*}

\subsection{Selection rules\label{sec: selection rules}}
The spin susceptibility measured in neutron scattering experiments obeys the two selection rules: (i) the allowed transitions must satisfy the condition $|\Delta S|\leq 1$, where $\Delta S$ is the difference  between the total spins of the final and  initial states; (ii) transitions between states both having zero total spin are forbidden.

\section{Magnetic response of compensated and uncompensated clusters\label{sec: results}}


\subsection{Compensated clusters}

We consider two compensated clusters with $N=12$ and $N=16$, see Fig \ref{fig:clusters}.  The numerically calculated magnetic responses of these two clusters are presented in Fig. \ref{fig:compensated}. Below we discuss their main features.

{\it Spin gap}. The ground state is a singlet separated from the first excited state by a gap. According to the  selection rule (ii), there are no transitions at zero frequency. As a consequence, spin response at low frequencies vanishes.

{\it Singlet-to-triplet excitations at $\omega_0\simeq J/2\simeq 60$ meV.} This is the most prominent feature seen in the plot of the integrated susceptibility $\chi''(\omega)$. It originates from the transition between the spin-singlet ground state and the lowest spin-1 excited state. The value of this frequency fits well the typical frequency $(40-70)$ meV of the broad peak  discussed in Sec. \ref{sec: experiments}.
In the momentum plane, the peak is localized around $\Q=(\pi,\pi)$ without any sign of splitting.


{\it Features of  $\chi''(\Q,\omega)$.}
The upward dispersion of  $\chi''(\Q,\omega)$ at frequencies  $\omega > \omega_0$  is present for  both clusters. The shapes of sections of  $\chi''(\Q,\omega)$ at a given $\omega$ are non-universal. The ring-shape pattern of $\chi''(\Q,\omega)$ seen in Fig. \ref{fig:compensated} for some of constant-$\omega$ sections  is often observed in the experiments.\cite{fujita2011progress,brooks2007handbook} When constant-$\omega$ sections are not ring-shaped, their orientations  depend on the orientations of the cluster's boundaries and often exhibit  $\pi/4$ rotations with changing $\omega$.


\medskip

We note that, as illustrated in Fig. \ref{fig:clusters}, the cluster with $N=12$ can approximate the basic building block of the spin-vortex checkerboard pattern proposed in Refs. \onlinecite{fine2007magnetic,dolgirev2017pseudogap,bhartiya2017superconductivity}. The coupling between these blocks is expected to be relatively small because of the smaller size of staggered spin polarizations at block's boundaries. The corners of these blocks are also separated from each other by spin vortex cores where the staggered spin polarizations almost vanish.

\subsection{Uncompensated clusters}

Three uncompensated clusters were considered, with $N=5$, $N=9$ and $N=13$, see Fig. \ref{fig:clusters}. According to the inequality (5) imposed by the Lieb-Mattis theorem the total spin of the ground states  of these clusters, $S_{gs}$, does not exceed $3/2$, $1/2$ and $5/2$, respectively. We find that, in fact, $S_{gs}$ assumes the above maximal values, in line with ref. \onlinecite{lieb1989two}.

We note here that the degeneracy of the ground state is taken into account in Eqs. \eqref{S(Q,w)} and \eqref{S(w)} by averaging over the ground state subspace. This procedure is equivalent to performing calculations at finite temperature $T$ and then taking the limit $T\rightarrow 0$.

The numerically obtained magnetic responses for uncompensated clusters are presented in Fig. \ref{fig:uncompensated}.   Their main features are the following:

{\it Response at $\omega\simeq 0$}. Since the transitions between the components of the ground state multiplet  are not forbidden by the selection rules, the magnetic response has a peak at $\omega=0$. The higher the multiplet degeneracy, the stronger the peak intensity. In the momentum plane, the intensity associated with this peak is localized around $\Q=(\pi,\pi)$ without splitting.

{\it Finite-frequency peaks.}  The lowest finite-frequency peaks for clusters with $N=5$ and $N=13$ are located in the frequency range (40-70) meV which is consistent with the experimentally observed broad peak discussed in Sec. \ref{sec: experiments}. At the same time, the lowest finite-frequency peak of the cluster with $N=9$ is located at a significantly higher frequency of about 100 meV.

{\it Dispersion of  $\chi''(\Q,\omega)$.} The dispersion of varying shapes around $\Q=(\pi,\pi)$ is generally present away from the zero frequency. As in the case of compensated clusters, the shape of the  sections of  $\chi''(\Q,\omega)$ at a given $\omega$  correlates with the orientation of the clusters.




\subsection{Spin response from the mixture of different clusters}

\begin{figure}[t]
\includegraphics[width = \textwidth]{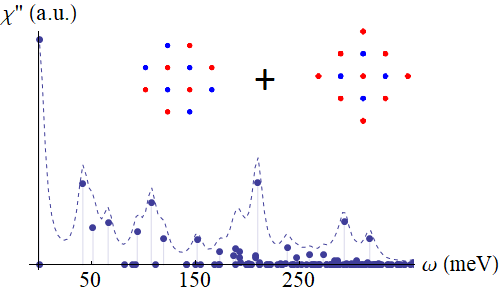}
 \caption{
Local susceptibility $\chi''(\omega)$ of the equally weighted mixture of clusters with $N=12$ and $N=13$. The Lorentzian broadening here is twice smaller than in the rest of the article, $\Gamma=0.05J=6$ meV.
 \label{fig:mixture}
 }
\end{figure}

It is possible that in a real system more than one type of clusters contribute to the magnetic response.
Magnetic response from a population of different clusters can be obtained by combining responses from individual clusters with corresponding weights.
As an illustration, we combine in Fig. \ref{fig:mixture} the responses from two clusters -- a compensated cluster with $N=12$ and uncompensated cluster with $N=13$.

\section{Summary and discussion\label{sec: conclusions}}

Our findings exhibited in Figs. \ref{fig:compensated}-\ref{fig:mixture} show that the magnetic responses from  finite clusters of spin $1/2$, despite high variability, exhibit features strongly reminiscent of the response measured in cuprates. These features are:\\
(i) spin gap in compensated clusters,\\
(ii) zero frequency peak in uncompensated clusters,\\
(iii) pronounced broad peak at $\omega=(40-70)$ meV present in both compensated and uncompensated clusters.\\

The spin gap in compensated clusters has a rather robust origin, namely, the transition from the spin-singlet ground state to the lowest spin-triplet excited state. Therefore, the interpretation of this gap in terms of compensated spin clusters  is quite realistic and competitive with other proposed interpretations, e.g. based on spin ladders.\cite{barnes1993excitation} The  zero frequency peak in uncompensated clusters emerging due to the transitions within the ground-state multiplet is even more remarkable, given that such a feature in underdoped lanthanum cuprates is particularly difficult to explain within all kinds of popular infinite-lattice models. Finally, a pronounced broad peak   at $\omega=(40-70)$ meV  emerges naturally as a finite size effect if one assumes that the linear size of clusters is about four lattice constants -- an assumption supported by experimental data as well as theoretical considerations. This further strengthens the merits of the cluster paradigm. We also note that all clusters exhibit upward dispersion of $\chi''(\Q,\omega)$ which is, however, typical also for a broad  class of infinite lattice models.\cite{tranquada1989neutron,hayden1996high,barnes1993excitation,vojta2004magnetic,uhrig2005magnetic,seibold2005magnetic,seibold2006doping,yao2006magnetic,james2012magnetic,andrade2012disorder}

At the same time, there are several features of spin response in cuprates which are not seen in clusters. These are: (i) the downward dispersion below the broad peak at  $\omega_0=(40-70)$ meV (lower part of the "hourglass"), and (ii) the sharp peak at low frequencies of about (7--18) meV seen in lanthanum cuprates.
The above discrepancies are not surprising  ---  a model of isolated clusters should not be expected to reproduce well all low-frequency features, because they can depend on intercluster interactions and on the interactions with itinerant charge carriers and phonons.\cite{egami2010spin-lattice,egami2010spin} We note, however, that the mercury family of hole-doped cuprates and the praseodymium family of electron-doped cuprates exhibit the "wine glass"~response\cite{chan2016commensurate} more consistent with our cluster calculations. Whether this is an indication of non-interacting clusters or a mere coincidence remains to be clarified.


To summarize, we analysed the  magnetic spin response in cuprates on the basis of the assumption that it may be coming from a collection of spin-1/2 clusters. We demonstrated that this approach is quite promising --- it provides simple physical interpretations for a number of common features of the cuprate magnetic response, including the spin gap and the zero frequency peak.

\begin{acknowledgments}
{\it Acknowledgements.}  This work was supported by Skoltech NGP
Program (Skoltech-MIT joint project).
\end{acknowledgments}

\bibliography{C:/D/Work/QM/Bibs/superconductivity}

\providecommand{\noopsort}[1]{}\providecommand{\singleletter}[1]{#1}%
\begin{thebibliography}{59}%
\makeatletter
\providecommand \@ifxundefined [1]{%
 \@ifx{#1\undefined}
}%
\providecommand \@ifnum [1]{%
 \ifnum #1\expandafter \@firstoftwo
 \else \expandafter \@secondoftwo
 \fi
}%
\providecommand \@ifx [1]{%
 \ifx #1\expandafter \@firstoftwo
 \else \expandafter \@secondoftwo
 \fi
}%
\providecommand \natexlab [1]{#1}%
\providecommand \enquote  [1]{``#1''}%
\providecommand \bibnamefont  [1]{#1}%
\providecommand \bibfnamefont [1]{#1}%
\providecommand \citenamefont [1]{#1}%
\providecommand \href@noop [0]{\@secondoftwo}%
\providecommand \href [0]{\begingroup \@sanitize@url \@href}%
\providecommand \@href[1]{\@@startlink{#1}\@@href}%
\providecommand \@@href[1]{\endgroup#1\@@endlink}%
\providecommand \@sanitize@url [0]{\catcode `\\12\catcode `\$12\catcode
  `\&12\catcode `\#12\catcode `\^12\catcode `\_12\catcode `\%12\relax}%
\providecommand \@@startlink[1]{}%
\providecommand \@@endlink[0]{}%
\providecommand \url  [0]{\begingroup\@sanitize@url \@url }%
\providecommand \@url [1]{\endgroup\@href {#1}{\urlprefix }}%
\providecommand \urlprefix  [0]{URL }%
\providecommand \Eprint [0]{\href }%
\providecommand \doibase [0]{http://dx.doi.org/}%
\providecommand \selectlanguage [0]{\@gobble}%
\providecommand \bibinfo  [0]{\@secondoftwo}%
\providecommand \bibfield  [0]{\@secondoftwo}%
\providecommand \translation [1]{[#1]}%
\providecommand \BibitemOpen [0]{}%
\providecommand \bibitemStop [0]{}%
\providecommand \bibitemNoStop [0]{.\EOS\space}%
\providecommand \EOS [0]{\spacefactor3000\relax}%
\providecommand \BibitemShut  [1]{\csname bibitem#1\endcsname}%
\let\auto@bib@innerbib\@empty
\bibitem [{\citenamefont {Fujita}\ \emph {et~al.}(2012)\citenamefont {Fujita},
  \citenamefont {Hiraka}, \citenamefont {Matsuda}, \citenamefont {Matsuura},
  \citenamefont {Tranquada}, \citenamefont {Wakimoto}, \citenamefont {Xu},\
  and\ \citenamefont {Yamada}}]{fujita2011progress}%
  \BibitemOpen
  \bibfield  {author} {\bibinfo {author} {\bibfnamefont {M.}~\bibnamefont
  {Fujita}}, \bibinfo {author} {\bibfnamefont {H.}~\bibnamefont {Hiraka}},
  \bibinfo {author} {\bibfnamefont {M.}~\bibnamefont {Matsuda}}, \bibinfo
  {author} {\bibfnamefont {M.}~\bibnamefont {Matsuura}}, \bibinfo {author}
  {\bibfnamefont {J.~M.}\ \bibnamefont {Tranquada}}, \bibinfo {author}
  {\bibfnamefont {S.}~\bibnamefont {Wakimoto}}, \bibinfo {author}
  {\bibfnamefont {G.}~\bibnamefont {Xu}}, \ and\ \bibinfo {author}
  {\bibfnamefont {K.}~\bibnamefont {Yamada}},\ }\bibfield  {title} {\enquote
  {\bibinfo {title} {Progress in neutron scattering studies of spin excitations
  in high-${T}_c$ cuprates},}\ }\href@noop {} {\bibfield  {journal} {\bibinfo
  {journal} {Journal of the Physical Society of Japan}\ }\textbf {\bibinfo
  {volume} {81}},\ \bibinfo {pages} {011007} (\bibinfo {year}
  {2012})}\BibitemShut {NoStop}%
\bibitem [{\citenamefont {Brooks}(2007)}]{brooks2007handbook}%
  \BibitemOpen
  \bibfield  {author} {\bibinfo {author} {\bibfnamefont {JS}~\bibnamefont
  {Brooks}},\ }\href@noop {} {\emph {\bibinfo {title} {Handbook of
  High-Temperature Superconductivity: Theory and Experiment}}}\ (\bibinfo
  {publisher} {Springer Science \& Business Media},\ \bibinfo {year}
  {2007})\BibitemShut {NoStop}%
\bibitem [{sup()}]{supplement}%
  \BibitemOpen
  \href@noop {} {}\bibinfo {note} {
  reproduce in the Supplemental Material several plots of spin response in
  cuprates obtained 
  Material in arXiv 1712.09979, where several plots of spin response in
  cuprates obtained in neutron scattering experiments are reproduced for the
  convenience of readers.}\BibitemShut {Stop}%
\bibitem [{\citenamefont {Miao}\ \emph {et~al.}(2017)\citenamefont {Miao},
  \citenamefont {Lorenzana}, \citenamefont {Seibold}, \citenamefont {Peng},
  \citenamefont {Amorese}, \citenamefont {Yakhou-Harris}, \citenamefont
  {Kummer}, \citenamefont {Brookes}, \citenamefont {Konik}, \citenamefont
  {Thampy} \emph {et~al.}}]{miao2017high}%
  \BibitemOpen
  \bibfield  {author} {\bibinfo {author} {\bibfnamefont {H}~\bibnamefont
  {Miao}}, \bibinfo {author} {\bibfnamefont {Jos{\'e}}\ \bibnamefont
  {Lorenzana}}, \bibinfo {author} {\bibfnamefont {G{\"o}tz}\ \bibnamefont
  {Seibold}}, \bibinfo {author} {\bibfnamefont {YY}~\bibnamefont {Peng}},
  \bibinfo {author} {\bibfnamefont {A}~\bibnamefont {Amorese}}, \bibinfo
  {author} {\bibfnamefont {F}~\bibnamefont {Yakhou-Harris}}, \bibinfo {author}
  {\bibfnamefont {K}~\bibnamefont {Kummer}}, \bibinfo {author} {\bibfnamefont
  {NB}~\bibnamefont {Brookes}}, \bibinfo {author} {\bibfnamefont
  {RM}~\bibnamefont {Konik}}, \bibinfo {author} {\bibfnamefont {V}~\bibnamefont
  {Thampy}},  \emph {et~al.},\ }\bibfield  {title} {\enquote {\bibinfo {title}
  {High-temperature charge density wave correlations in
  {L}a$_{1.875}${B}a$_{0.125}${C}u{O}$_4$ without spin--charge locking},}\
  }\href@noop {} {\bibfield  {journal} {\bibinfo  {journal} {Proceedings of the
  National Academy of Sciences}\ ,\ \bibinfo {pages} {201708549}} (\bibinfo
  {year} {2017})}\BibitemShut {NoStop}%
\bibitem [{\citenamefont {Seibold}\ and\ \citenamefont
  {Lorenzana}(2005)}]{seibold2005magnetic}%
  \BibitemOpen
  \bibfield  {author} {\bibinfo {author} {\bibfnamefont {G{\"o}tz}\
  \bibnamefont {Seibold}}\ and\ \bibinfo {author} {\bibfnamefont {Jose}\
  \bibnamefont {Lorenzana}},\ }\bibfield  {title} {\enquote {\bibinfo {title}
  {Magnetic fluctuations of stripes in the high temperature cuprate
  superconductors},}\ }\href@noop {} {\bibfield  {journal} {\bibinfo  {journal}
  {Physical review letters}\ }\textbf {\bibinfo {volume} {94}},\ \bibinfo
  {pages} {107006} (\bibinfo {year} {2005})}\BibitemShut {NoStop}%
\bibitem [{\citenamefont {Seibold}\ and\ \citenamefont
  {Lorenzana}(2006)}]{seibold2006doping}%
  \BibitemOpen
  \bibfield  {author} {\bibinfo {author} {\bibfnamefont {G{\"o}tz}\
  \bibnamefont {Seibold}}\ and\ \bibinfo {author} {\bibfnamefont {Jose}\
  \bibnamefont {Lorenzana}},\ }\bibfield  {title} {\enquote {\bibinfo {title}
  {Doping dependence of spin excitations in the stripe phase of high-t c
  superconductors},}\ }\href@noop {} {\bibfield  {journal} {\bibinfo  {journal}
  {Physical Review B}\ }\textbf {\bibinfo {volume} {73}},\ \bibinfo {pages}
  {144515} (\bibinfo {year} {2006})}\BibitemShut {NoStop}%
\bibitem [{\citenamefont {Kung}\ \emph {et~al.}(2015)\citenamefont {Kung},
  \citenamefont {Nowadnick}, \citenamefont {Jia}, \citenamefont {Johnston},
  \citenamefont {Moritz}, \citenamefont {Scalettar},\ and\ \citenamefont
  {Devereaux}}]{kung2015doping}%
  \BibitemOpen
  \bibfield  {author} {\bibinfo {author} {\bibfnamefont {Y.~F.}\ \bibnamefont
  {Kung}}, \bibinfo {author} {\bibfnamefont {E.~A.}\ \bibnamefont {Nowadnick}},
  \bibinfo {author} {\bibfnamefont {C.~J.}\ \bibnamefont {Jia}}, \bibinfo
  {author} {\bibfnamefont {S.}~\bibnamefont {Johnston}}, \bibinfo {author}
  {\bibfnamefont {B.}~\bibnamefont {Moritz}}, \bibinfo {author} {\bibfnamefont
  {R.~T.}\ \bibnamefont {Scalettar}}, \ and\ \bibinfo {author} {\bibfnamefont
  {T.~P.}\ \bibnamefont {Devereaux}},\ }\bibfield  {title} {\enquote {\bibinfo
  {title} {Doping evolution of spin and charge excitations in the hubbard
  model},}\ }\href {\doibase 10.1103/PhysRevB.92.195108} {\bibfield  {journal}
  {\bibinfo  {journal} {Phys. Rev. B}\ }\textbf {\bibinfo {volume} {92}},\
  \bibinfo {pages} {195108} (\bibinfo {year} {2015})}\BibitemShut {NoStop}%
\bibitem [{\citenamefont {Onufrieva}\ and\ \citenamefont
  {Pfeuty}(2002)}]{onufrieva2002spin}%
  \BibitemOpen
  \bibfield  {author} {\bibinfo {author} {\bibfnamefont {F.}~\bibnamefont
  {Onufrieva}}\ and\ \bibinfo {author} {\bibfnamefont {P.}~\bibnamefont
  {Pfeuty}},\ }\bibfield  {title} {\enquote {\bibinfo {title} {Spin dynamics of
  a two-dimensional metal in a superconducting state: Application to the
  high-${T}_{c}$ cuprates},}\ }\href {\doibase 10.1103/PhysRevB.65.054515}
  {\bibfield  {journal} {\bibinfo  {journal} {Phys. Rev. B}\ }\textbf {\bibinfo
  {volume} {65}},\ \bibinfo {pages} {054515} (\bibinfo {year}
  {2002})}\BibitemShut {NoStop}%
\bibitem [{\citenamefont {Sushkov}\ and\ \citenamefont
  {Kotov}(2005)}]{sushkov2005theory}%
  \BibitemOpen
  \bibfield  {author} {\bibinfo {author} {\bibfnamefont {Oleg~P.}\ \bibnamefont
  {Sushkov}}\ and\ \bibinfo {author} {\bibfnamefont {Valeri~N.}\ \bibnamefont
  {Kotov}},\ }\bibfield  {title} {\enquote {\bibinfo {title} {Theory of
  incommensurate magnetic correlations across the insulator-superconductor
  transition of underdoped {L}a$_{2-x}${S}r$_x${C}u{O}$_4$},}\ }\href {\doibase
  10.1103/PhysRevLett.94.097005} {\bibfield  {journal} {\bibinfo  {journal}
  {Phys. Rev. Lett.}\ }\textbf {\bibinfo {volume} {94}},\ \bibinfo {pages}
  {097005} (\bibinfo {year} {2005})}\BibitemShut {NoStop}%
\bibitem [{\citenamefont {James}\ \emph {et~al.}(2012)\citenamefont {James},
  \citenamefont {Konik},\ and\ \citenamefont {Rice}}]{james2012magnetic}%
  \BibitemOpen
  \bibfield  {author} {\bibinfo {author} {\bibfnamefont {A.~J.~A.}\
  \bibnamefont {James}}, \bibinfo {author} {\bibfnamefont {R.~M.}\ \bibnamefont
  {Konik}}, \ and\ \bibinfo {author} {\bibfnamefont {T.~M.}\ \bibnamefont
  {Rice}},\ }\bibfield  {title} {\enquote {\bibinfo {title} {Magnetic response
  in the underdoped cuprates},}\ }\href {\doibase 10.1103/PhysRevB.86.100508}
  {\bibfield  {journal} {\bibinfo  {journal} {Phys. Rev. B}\ }\textbf {\bibinfo
  {volume} {86}},\ \bibinfo {pages} {100508} (\bibinfo {year}
  {2012})}\BibitemShut {NoStop}%
\bibitem [{\citenamefont {Eremin}\ \emph {et~al.}(2012)\citenamefont {Eremin},
  \citenamefont {Shigapov},\ and\ \citenamefont {Eremin}}]{eremin2012dual}%
  \BibitemOpen
  \bibfield  {author} {\bibinfo {author} {\bibfnamefont {MV}~\bibnamefont
  {Eremin}}, \bibinfo {author} {\bibfnamefont {IM}~\bibnamefont {Shigapov}}, \
  and\ \bibinfo {author} {\bibfnamefont {IM}~\bibnamefont {Eremin}},\
  }\bibfield  {title} {\enquote {\bibinfo {title} {Dual features of magnetic
  susceptibility in superconducting cuprates: a comparison to inelastic neutron
  scattering},}\ }\href@noop {} {\bibfield  {journal} {\bibinfo  {journal} {The
  European Physical Journal B}\ }\textbf {\bibinfo {volume} {85}},\ \bibinfo
  {pages} {131} (\bibinfo {year} {2012})}\BibitemShut {NoStop}%
\bibitem [{\citenamefont {S.~Uhrig}\ \emph {et~al.}(2005)\citenamefont
  {S.~Uhrig}, \citenamefont {P.~Schmidt},\ and\ \citenamefont
  {Gr{\"u}ninger}}]{uhrig2005magnetic}%
  \BibitemOpen
  \bibfield  {author} {\bibinfo {author} {\bibfnamefont {G}~\bibnamefont
  {S.~Uhrig}}, \bibinfo {author} {\bibfnamefont {K}~\bibnamefont {P.~Schmidt}},
  \ and\ \bibinfo {author} {\bibfnamefont {M}~\bibnamefont {Gr{\"u}ninger}},\
  }\bibfield  {title} {\enquote {\bibinfo {title} {Magnetic excitations in
  bilayer high-temperature superconductors with stripe correlations},}\
  }\href@noop {} {\bibfield  {journal} {\bibinfo  {journal} {Journal of the
  Physical Society of Japan}\ }\textbf {\bibinfo {volume} {74}},\ \bibinfo
  {pages} {86--97} (\bibinfo {year} {2005})}\BibitemShut {NoStop}%
\bibitem [{\citenamefont {Vojta}\ and\ \citenamefont
  {Ulbricht}(2004)}]{vojta2004magnetic}%
  \BibitemOpen
  \bibfield  {author} {\bibinfo {author} {\bibfnamefont {Matthias}\
  \bibnamefont {Vojta}}\ and\ \bibinfo {author} {\bibfnamefont {Tobias}\
  \bibnamefont {Ulbricht}},\ }\bibfield  {title} {\enquote {\bibinfo {title}
  {Magnetic excitations in a bond-centered stripe phase: Spin waves far from
  the semiclassical limit},}\ }\href@noop {} {\bibfield  {journal} {\bibinfo
  {journal} {Phys. Rev. Lett.}\ }\textbf {\bibinfo {volume} {93}},\ \bibinfo
  {pages} {127002} (\bibinfo {year} {2004})}\BibitemShut {NoStop}%
\bibitem [{\citenamefont {Yao}\ \emph {et~al.}(2006)\citenamefont {Yao},
  \citenamefont {Carlson},\ and\ \citenamefont {Campbell}}]{yao2006magnetic}%
  \BibitemOpen
  \bibfield  {author} {\bibinfo {author} {\bibfnamefont {DX}~\bibnamefont
  {Yao}}, \bibinfo {author} {\bibfnamefont {EW}~\bibnamefont {Carlson}}, \ and\
  \bibinfo {author} {\bibfnamefont {DK}~\bibnamefont {Campbell}},\ }\bibfield
  {title} {\enquote {\bibinfo {title} {Magnetic excitations of stripes and
  checkerboards in the cuprates},}\ }\href@noop {} {\bibfield  {journal}
  {\bibinfo  {journal} {Physical Review B}\ }\textbf {\bibinfo {volume} {73}},\
  \bibinfo {pages} {224525} (\bibinfo {year} {2006})}\BibitemShut {NoStop}%
\bibitem [{\citenamefont {Mang}\ \emph {et~al.}(2004)\citenamefont {Mang},
  \citenamefont {Vajk}, \citenamefont {Arvanitaki}, \citenamefont {Lynn},\ and\
  \citenamefont {Greven}}]{mang2004spin}%
  \BibitemOpen
  \bibfield  {author} {\bibinfo {author} {\bibfnamefont {P.~K.}\ \bibnamefont
  {Mang}}, \bibinfo {author} {\bibfnamefont {O.~P.}\ \bibnamefont {Vajk}},
  \bibinfo {author} {\bibfnamefont {A.}~\bibnamefont {Arvanitaki}}, \bibinfo
  {author} {\bibfnamefont {J.~W.}\ \bibnamefont {Lynn}}, \ and\ \bibinfo
  {author} {\bibfnamefont {M.}~\bibnamefont {Greven}},\ }\bibfield  {title}
  {\enquote {\bibinfo {title} {Spin correlations and magnetic order in
  nonsuperconducting {N}d$_{2- x}${C}e$_x${C}u{O}$_{4\pm\delta}$},}\
  }\href@noop {} {\bibfield  {journal} {\bibinfo  {journal} {Phys. Rev. Lett.}\
  }\textbf {\bibinfo {volume} {93}},\ \bibinfo {pages} {027002} (\bibinfo
  {year} {2004})}\BibitemShut {NoStop}%
\bibitem [{\citenamefont {Andrade}\ and\ \citenamefont
  {Vojta}(2012)}]{andrade2012disorder}%
  \BibitemOpen
  \bibfield  {author} {\bibinfo {author} {\bibfnamefont {Eric~C.}\ \bibnamefont
  {Andrade}}\ and\ \bibinfo {author} {\bibfnamefont {Matthias}\ \bibnamefont
  {Vojta}},\ }\bibfield  {title} {\enquote {\bibinfo {title} {Disorder, cluster
  spin glass, and hourglass spectra in striped magnetic insulators},}\ }\href
  {\doibase 10.1103/PhysRevLett.109.147201} {\bibfield  {journal} {\bibinfo
  {journal} {Phys. Rev. Lett.}\ }\textbf {\bibinfo {volume} {109}},\ \bibinfo
  {pages} {147201} (\bibinfo {year} {2012})}\BibitemShut {NoStop}%
\bibitem [{\citenamefont {Abanov}\ and\ \citenamefont
  {Chubukov}(1999)}]{abanov1999relation}%
  \BibitemOpen
  \bibfield  {author} {\bibinfo {author} {\bibfnamefont {Ar.}\ \bibnamefont
  {Abanov}}\ and\ \bibinfo {author} {\bibfnamefont {Andrey~V.}\ \bibnamefont
  {Chubukov}},\ }\bibfield  {title} {\enquote {\bibinfo {title} {A relation
  between the resonance neutron peak and arpes data in cuprates},}\ }\href
  {\doibase 10.1103/PhysRevLett.83.1652} {\bibfield  {journal} {\bibinfo
  {journal} {Phys. Rev. Lett.}\ }\textbf {\bibinfo {volume} {83}},\ \bibinfo
  {pages} {1652--1655} (\bibinfo {year} {1999})}\BibitemShut {NoStop}%
\bibitem [{\citenamefont {Bianconi}(2013)}]{bianconi2013quantum}%
  \BibitemOpen
  \bibfield  {author} {\bibinfo {author} {\bibfnamefont {Antonio}\ \bibnamefont
  {Bianconi}},\ }\bibfield  {title} {\enquote {\bibinfo {title} {Quantum
  materials: Shape resonances in superstripes},}\ }\href@noop {} {\bibfield
  {journal} {\bibinfo  {journal} {Nature Physics}\ }\textbf {\bibinfo {volume}
  {9}},\ \bibinfo {pages} {536--537} (\bibinfo {year} {2013})}\BibitemShut
  {NoStop}%
\bibitem [{\citenamefont {Campi}\ \emph {et~al.}(2015)\citenamefont {Campi},
  \citenamefont {Bianconi}, \citenamefont {Poccia}, \citenamefont {Bianconi},
  \citenamefont {Barba}, \citenamefont {Arrighetti}, \citenamefont {Innocenti},
  \citenamefont {Karpinski}, \citenamefont {Zhigadlo}, \citenamefont {Kazakov}
  \emph {et~al.}}]{campi2015inhomogeneity}%
  \BibitemOpen
  \bibfield  {author} {\bibinfo {author} {\bibfnamefont {Gaetano}\ \bibnamefont
  {Campi}}, \bibinfo {author} {\bibfnamefont {Antonio}\ \bibnamefont
  {Bianconi}}, \bibinfo {author} {\bibfnamefont {Nicola}\ \bibnamefont
  {Poccia}}, \bibinfo {author} {\bibfnamefont {Ginestra}\ \bibnamefont
  {Bianconi}}, \bibinfo {author} {\bibfnamefont {L}~\bibnamefont {Barba}},
  \bibinfo {author} {\bibfnamefont {Gianmichele}\ \bibnamefont {Arrighetti}},
  \bibinfo {author} {\bibfnamefont {Davide}\ \bibnamefont {Innocenti}},
  \bibinfo {author} {\bibfnamefont {J}~\bibnamefont {Karpinski}}, \bibinfo
  {author} {\bibfnamefont {Nikolai~D}\ \bibnamefont {Zhigadlo}}, \bibinfo
  {author} {\bibfnamefont {Sergey~M}\ \bibnamefont {Kazakov}},  \emph
  {et~al.},\ }\bibfield  {title} {\enquote {\bibinfo {title} {Inhomogeneity of
  charge-density-wave order and quenched disorder in a high-tc
  superconductor},}\ }\href@noop {} {\bibfield  {journal} {\bibinfo  {journal}
  {Nature}\ }\textbf {\bibinfo {volume} {525}},\ \bibinfo {pages} {359--362}
  (\bibinfo {year} {2015})}\BibitemShut {NoStop}%
\bibitem [{\citenamefont {Emery}\ and\ \citenamefont
  {Kivelson}(1993)}]{emery1993frustrated}%
  \BibitemOpen
  \bibfield  {author} {\bibinfo {author} {\bibfnamefont {Victor~J}\
  \bibnamefont {Emery}}\ and\ \bibinfo {author} {\bibfnamefont
  {SA}~\bibnamefont {Kivelson}},\ }\bibfield  {title} {\enquote {\bibinfo
  {title} {Frustrated electronic phase separation and high-temperature
  superconductors},}\ }\href@noop {} {\bibfield  {journal} {\bibinfo  {journal}
  {Physica C: Superconductivity}\ }\textbf {\bibinfo {volume} {209}},\ \bibinfo
  {pages} {597--621} (\bibinfo {year} {1993})}\BibitemShut {NoStop}%
\bibitem [{\citenamefont {Fine}\ and\ \citenamefont
  {Egami}(2008)}]{fine2008phase}%
  \BibitemOpen
  \bibfield  {author} {\bibinfo {author} {\bibfnamefont {B.~V.}\ \bibnamefont
  {Fine}}\ and\ \bibinfo {author} {\bibfnamefont {T.}~\bibnamefont {Egami}},\
  }\bibfield  {title} {\enquote {\bibinfo {title} {Phase separation in the
  vicinity of quantum-critical doping concentration: Implications for
  high-temperature superconductors},}\ }\href {\doibase
  10.1103/PhysRevB.77.014519} {\bibfield  {journal} {\bibinfo  {journal} {Phys.
  Rev. B}\ }\textbf {\bibinfo {volume} {77}},\ \bibinfo {pages} {014519}
  (\bibinfo {year} {2008})}\BibitemShut {NoStop}%
\bibitem [{\citenamefont {Fine}(2007)}]{fine2007magnetic}%
  \BibitemOpen
  \bibfield  {author} {\bibinfo {author} {\bibfnamefont {Boris~V.}\
  \bibnamefont {Fine}},\ }\bibfield  {title} {\enquote {\bibinfo {title}
  {Magnetic vortices instead of stripes: Another interpretation of magnetic
  neutron scattering in lanthanum cuprates},}\ }\href {\doibase
  10.1103/PhysRevB.75.060504} {\bibfield  {journal} {\bibinfo  {journal} {Phys.
  Rev. B}\ }\textbf {\bibinfo {volume} {75}},\ \bibinfo {pages} {060504}
  (\bibinfo {year} {2007})}\BibitemShut {NoStop}%
\bibitem [{\citenamefont {Dolgirev}\ and\ \citenamefont
  {Fine}(2017)}]{dolgirev2017pseudogap}%
  \BibitemOpen
  \bibfield  {author} {\bibinfo {author} {\bibfnamefont {Pavel~E.}\
  \bibnamefont {Dolgirev}}\ and\ \bibinfo {author} {\bibfnamefont {Boris~V.}\
  \bibnamefont {Fine}},\ }\bibfield  {title} {\enquote {\bibinfo {title}
  {Pseudogap and fermi surface in the presence of a spin-vortex checkerboard
  for 1/8-doped lanthanum cuprates},}\ }\href {\doibase
  10.1103/PhysRevB.96.075137} {\bibfield  {journal} {\bibinfo  {journal} {Phys.
  Rev. B}\ }\textbf {\bibinfo {volume} {96}},\ \bibinfo {pages} {075137}
  (\bibinfo {year} {2017})}\BibitemShut {NoStop}%
\bibitem [{\citenamefont {Bhartiya}\ and\ \citenamefont
  {Fine}(2017)}]{bhartiya2017superconductivity}%
  \BibitemOpen
  \bibfield  {author} {\bibinfo {author} {\bibfnamefont {Vivek~K}\ \bibnamefont
  {Bhartiya}}\ and\ \bibinfo {author} {\bibfnamefont {Boris~V}\ \bibnamefont
  {Fine}},\ }\bibfield  {title} {\enquote {\bibinfo {title} {Superconductivity
  model for a spin-vortex checkerboard},}\ }\href@noop {} {\bibfield  {journal}
  {\bibinfo  {journal} {arXiv preprint arXiv:1703.09979}\ } (\bibinfo {year}
  {2017})}\BibitemShut {NoStop}%
\bibitem [{\citenamefont {Tranquada}\ \emph {et~al.}(2004)\citenamefont
  {Tranquada}, \citenamefont {Woo}, \citenamefont {Perring}, \citenamefont
  {Goka}, \citenamefont {Gu}, \citenamefont {Xu}, \citenamefont {Fujita},\ and\
  \citenamefont {Yamada}}]{tranquada2004quantum}%
  \BibitemOpen
  \bibfield  {author} {\bibinfo {author} {\bibfnamefont {J.~M.}\ \bibnamefont
  {Tranquada}}, \bibinfo {author} {\bibfnamefont {H.}~\bibnamefont {Woo}},
  \bibinfo {author} {\bibfnamefont {T.~G.}\ \bibnamefont {Perring}}, \bibinfo
  {author} {\bibfnamefont {H.}~\bibnamefont {Goka}}, \bibinfo {author}
  {\bibfnamefont {G.~D.}\ \bibnamefont {Gu}}, \bibinfo {author} {\bibfnamefont
  {G.}~\bibnamefont {Xu}}, \bibinfo {author} {\bibfnamefont {M.}~\bibnamefont
  {Fujita}}, \ and\ \bibinfo {author} {\bibfnamefont {K.}~\bibnamefont
  {Yamada}},\ }\bibfield  {title} {\enquote {\bibinfo {title} {Quantum magnetic
  excitations from stripes in copper oxide superconductors},}\ }\href@noop {}
  {\bibfield  {journal} {\bibinfo  {journal} {Nature}\ }\textbf {\bibinfo
  {volume} {429}},\ \bibinfo {pages} {534--538} (\bibinfo {year}
  {2004})}\BibitemShut {NoStop}%
\bibitem [{\citenamefont {Lipscombe}\ \emph {et~al.}(2009)\citenamefont
  {Lipscombe}, \citenamefont {Vignolle}, \citenamefont {Perring}, \citenamefont
  {Frost},\ and\ \citenamefont {Hayden}}]{lipscombe2009emergence}%
  \BibitemOpen
  \bibfield  {author} {\bibinfo {author} {\bibfnamefont {O.J.}\ \bibnamefont
  {Lipscombe}}, \bibinfo {author} {\bibfnamefont {B.}~\bibnamefont {Vignolle}},
  \bibinfo {author} {\bibfnamefont {T.G.}\ \bibnamefont {Perring}}, \bibinfo
  {author} {\bibfnamefont {C.D.}\ \bibnamefont {Frost}}, \ and\ \bibinfo
  {author} {\bibfnamefont {S.M.}\ \bibnamefont {Hayden}},\ }\bibfield  {title}
  {\enquote {\bibinfo {title} {Emergence of coherent magnetic excitations in
  the high temperature underdoped {L}a$_{2-x}${S}r$_x${C}uo$_4$ superconductor
  at low temperatures},}\ }\href@noop {} {\bibfield  {journal} {\bibinfo
  {journal} {Phys. Rev. Lett.}\ }\textbf {\bibinfo {volume} {102}},\ \bibinfo
  {pages} {167002} (\bibinfo {year} {2009})}\BibitemShut {NoStop}%
\bibitem [{\citenamefont {Vignolle}\ \emph {et~al.}(2007)\citenamefont
  {Vignolle}, \citenamefont {Hayden}, \citenamefont {McMorrow}, \citenamefont
  {R{\o}nnow}, \citenamefont {Lake}, \citenamefont {Frost},\ and\ \citenamefont
  {Perring}}]{vignolle2007two}%
  \BibitemOpen
  \bibfield  {author} {\bibinfo {author} {\bibfnamefont {B.}~\bibnamefont
  {Vignolle}}, \bibinfo {author} {\bibfnamefont {S.~M.}\ \bibnamefont
  {Hayden}}, \bibinfo {author} {\bibfnamefont {D.~F.}\ \bibnamefont
  {McMorrow}}, \bibinfo {author} {\bibfnamefont {H.~M.}\ \bibnamefont
  {R{\o}nnow}}, \bibinfo {author} {\bibfnamefont {B.}~\bibnamefont {Lake}},
  \bibinfo {author} {\bibfnamefont {C.~D.}\ \bibnamefont {Frost}}, \ and\
  \bibinfo {author} {\bibfnamefont {T.~G.}\ \bibnamefont {Perring}},\
  }\bibfield  {title} {\enquote {\bibinfo {title} {Two energy scales in the
  spin excitations of the high-temperature superconductor
  {L}a$_{2-x}${S}r$_x${C}uo$_4$},}\ }\href@noop {} {\bibfield  {journal}
  {\bibinfo  {journal} {Nature Physics}\ }\textbf {\bibinfo {volume} {3}},\
  \bibinfo {pages} {163--167} (\bibinfo {year} {2007})}\BibitemShut {NoStop}%
\bibitem [{\citenamefont {Lipscombe}\ \emph {et~al.}(2007)\citenamefont
  {Lipscombe}, \citenamefont {Hayden}, \citenamefont {Vignolle}, \citenamefont
  {McMorrow},\ and\ \citenamefont {Perring}}]{lipscombe2007persistence}%
  \BibitemOpen
  \bibfield  {author} {\bibinfo {author} {\bibfnamefont {O.~J.}\ \bibnamefont
  {Lipscombe}}, \bibinfo {author} {\bibfnamefont {S.~M.}\ \bibnamefont
  {Hayden}}, \bibinfo {author} {\bibfnamefont {B.}~\bibnamefont {Vignolle}},
  \bibinfo {author} {\bibfnamefont {D.~F.}\ \bibnamefont {McMorrow}}, \ and\
  \bibinfo {author} {\bibfnamefont {T.~G.}\ \bibnamefont {Perring}},\
  }\bibfield  {title} {\enquote {\bibinfo {title} {Persistence of
  high-frequency spin fluctuations in overdoped superconducting
  {L}a$_{2-x}${S}r$_x${C}uo$_4$ (x= 0.22)},}\ }\href@noop {} {\bibfield
  {journal} {\bibinfo  {journal} {Physical {R}eview {L}etters}\ }\textbf
  {\bibinfo {volume} {99}},\ \bibinfo {pages} {067002} (\bibinfo {year}
  {2007})}\BibitemShut {NoStop}%
\bibitem [{\citenamefont {Dai}\ \emph {et~al.}(2001)\citenamefont {Dai},
  \citenamefont {Mook}, \citenamefont {Hunt},\ and\ \citenamefont
  {Do\ifmmode~\breve{g}\else \u{g}\fi{}an}}]{dai2001evolution}%
  \BibitemOpen
  \bibfield  {author} {\bibinfo {author} {\bibfnamefont {Pengcheng}\
  \bibnamefont {Dai}}, \bibinfo {author} {\bibfnamefont {H.~A.}\ \bibnamefont
  {Mook}}, \bibinfo {author} {\bibfnamefont {R.~D.}\ \bibnamefont {Hunt}}, \
  and\ \bibinfo {author} {\bibfnamefont {F.}~\bibnamefont
  {Do\ifmmode~\breve{g}\else \u{g}\fi{}an}},\ }\bibfield  {title} {\enquote
  {\bibinfo {title} {Evolution of the resonance and incommensurate spin
  fluctuations in superconducting
  ${\mathrm{{y}ba}}_{2}{\mathrm{cu}}_{3}{\mathrm{o}}_{6+x}$},}\ }\href
  {\doibase 10.1103/PhysRevB.63.054525} {\bibfield  {journal} {\bibinfo
  {journal} {Phys. Rev. B}\ }\textbf {\bibinfo {volume} {63}},\ \bibinfo
  {pages} {054525} (\bibinfo {year} {2001})}\BibitemShut {NoStop}%
\bibitem [{\citenamefont {Chan}\ \emph
  {et~al.}(2016{\natexlab{a}})\citenamefont {Chan}, \citenamefont {Tang},
  \citenamefont {Dorow}, \citenamefont {Jeong}, \citenamefont {Mangin-Thro},
  \citenamefont {Veit}, \citenamefont {Ge}, \citenamefont {Abernathy},
  \citenamefont {Sidis}, \citenamefont {Bourges} \emph
  {et~al.}}]{chan2016hourglass}%
  \BibitemOpen
  \bibfield  {author} {\bibinfo {author} {\bibfnamefont {M.K.}\ \bibnamefont
  {Chan}}, \bibinfo {author} {\bibfnamefont {Y.}~\bibnamefont {Tang}}, \bibinfo
  {author} {\bibfnamefont {C.J.}\ \bibnamefont {Dorow}}, \bibinfo {author}
  {\bibfnamefont {J.}~\bibnamefont {Jeong}}, \bibinfo {author} {\bibfnamefont
  {L.}~\bibnamefont {Mangin-Thro}}, \bibinfo {author} {\bibfnamefont {M.J.}\
  \bibnamefont {Veit}}, \bibinfo {author} {\bibfnamefont {Y.}~\bibnamefont
  {Ge}}, \bibinfo {author} {\bibfnamefont {D.L.}\ \bibnamefont {Abernathy}},
  \bibinfo {author} {\bibfnamefont {Y.}~\bibnamefont {Sidis}}, \bibinfo
  {author} {\bibfnamefont {Ph.}\ \bibnamefont {Bourges}},  \emph {et~al.},\
  }\bibfield  {title} {\enquote {\bibinfo {title} {Hourglass dispersion and
  resonance of magnetic excitations in the superconducting state of the
  single-layer cuprate {H}g{B}a$_2${C}u{O}$_{4+\delta}$ near optimal doping},}\
  }\href@noop {} {\bibfield  {journal} {\bibinfo  {journal} {Phys. Rev. Lett.}\
  }\textbf {\bibinfo {volume} {117}},\ \bibinfo {pages} {277002} (\bibinfo
  {year} {2016}{\natexlab{a}})}\BibitemShut {NoStop}%
\bibitem [{\citenamefont {Chan}\ \emph
  {et~al.}(2016{\natexlab{b}})\citenamefont {Chan}, \citenamefont {Dorow},
  \citenamefont {Mangin-Thro}, \citenamefont {Tang}, \citenamefont {Ge},
  \citenamefont {Veit}, \citenamefont {Yu}, \citenamefont {Zhao}, \citenamefont
  {Christianson}, \citenamefont {Park} \emph {et~al.}}]{chan2016commensurate}%
  \BibitemOpen
  \bibfield  {author} {\bibinfo {author} {\bibfnamefont {M.K.}\ \bibnamefont
  {Chan}}, \bibinfo {author} {\bibfnamefont {C.J.}\ \bibnamefont {Dorow}},
  \bibinfo {author} {\bibfnamefont {L.}~\bibnamefont {Mangin-Thro}}, \bibinfo
  {author} {\bibfnamefont {Y.}~\bibnamefont {Tang}}, \bibinfo {author}
  {\bibfnamefont {Y.}~\bibnamefont {Ge}}, \bibinfo {author} {\bibfnamefont
  {M.J.}\ \bibnamefont {Veit}}, \bibinfo {author} {\bibfnamefont
  {G.}~\bibnamefont {Yu}}, \bibinfo {author} {\bibfnamefont {X.}~\bibnamefont
  {Zhao}}, \bibinfo {author} {\bibfnamefont {A.D.}\ \bibnamefont
  {Christianson}}, \bibinfo {author} {\bibfnamefont {J.T.}\ \bibnamefont
  {Park}},  \emph {et~al.},\ }\bibfield  {title} {\enquote {\bibinfo {title}
  {Commensurate antiferromagnetic excitations as a signature of the pseudogap
  in the tetragonal high-tc cuprate {H}g{B}a$_2${C}u{O}$_{4+\delta}$},}\
  }\href@noop {} {\bibfield  {journal} {\bibinfo  {journal} {Nat. Commun.}\
  }\textbf {\bibinfo {volume} {7}},\ \bibinfo {pages} {12244} (\bibinfo {year}
  {2016}{\natexlab{b}})}\BibitemShut {NoStop}%
\bibitem [{\citenamefont {Stock}\ \emph {et~al.}(2010)\citenamefont {Stock},
  \citenamefont {Cowley}, \citenamefont {Buyers}, \citenamefont {Frost},
  \citenamefont {Taylor}, \citenamefont {Peets}, \citenamefont {Liang},
  \citenamefont {Bonn},\ and\ \citenamefont {Hardy}}]{stock2010effect}%
  \BibitemOpen
  \bibfield  {author} {\bibinfo {author} {\bibfnamefont {C.}~\bibnamefont
  {Stock}}, \bibinfo {author} {\bibfnamefont {R.A.}\ \bibnamefont {Cowley}},
  \bibinfo {author} {\bibfnamefont {W.J.L.}\ \bibnamefont {Buyers}}, \bibinfo
  {author} {\bibfnamefont {C.D.}\ \bibnamefont {Frost}}, \bibinfo {author}
  {\bibfnamefont {J.W.}\ \bibnamefont {Taylor}}, \bibinfo {author}
  {\bibfnamefont {D.}~\bibnamefont {Peets}}, \bibinfo {author} {\bibfnamefont
  {R.}~\bibnamefont {Liang}}, \bibinfo {author} {\bibfnamefont
  {D.}~\bibnamefont {Bonn}}, \ and\ \bibinfo {author} {\bibfnamefont {W.N.}\
  \bibnamefont {Hardy}},\ }\bibfield  {title} {\enquote {\bibinfo {title}
  {Effect of the pseudogap on suppressing high energy inelastic neutron
  scattering in superconducting {YB}a$_2${C}u$_3${O}$_{6.5}$},}\ }\href@noop {}
  {\bibfield  {journal} {\bibinfo  {journal} {Physical Review B}\ }\textbf
  {\bibinfo {volume} {82}},\ \bibinfo {pages} {174505} (\bibinfo {year}
  {2010})}\BibitemShut {NoStop}%
\bibitem [{\citenamefont {Xu}\ \emph {et~al.}(2009)\citenamefont {Xu},
  \citenamefont {Gu}, \citenamefont {H{\"u}cker}, \citenamefont {Fauqu{\'e}},
  \citenamefont {Perring}, \citenamefont {Regnault},\ and\ \citenamefont
  {Tranquada}}]{xu2009testing}%
  \BibitemOpen
  \bibfield  {author} {\bibinfo {author} {\bibfnamefont {G.}~\bibnamefont
  {Xu}}, \bibinfo {author} {\bibfnamefont {G.D.}\ \bibnamefont {Gu}}, \bibinfo
  {author} {\bibfnamefont {M.}~\bibnamefont {H{\"u}cker}}, \bibinfo {author}
  {\bibfnamefont {B.}~\bibnamefont {Fauqu{\'e}}}, \bibinfo {author}
  {\bibfnamefont {T.G.}\ \bibnamefont {Perring}}, \bibinfo {author}
  {\bibfnamefont {L.P.}\ \bibnamefont {Regnault}}, \ and\ \bibinfo {author}
  {\bibfnamefont {J.M.}\ \bibnamefont {Tranquada}},\ }\bibfield  {title}
  {\enquote {\bibinfo {title} {Testing the itinerancy of spin dynamics in
  superconducting {Bi$_2$Sr$_2$CaCu$_2$O$_{8+\delta}$}},}\ }\href@noop {}
  {\bibfield  {journal} {\bibinfo  {journal} {Nature Physics}\ }\textbf
  {\bibinfo {volume} {5}},\ \bibinfo {pages} {642--646} (\bibinfo {year}
  {2009})}\BibitemShut {NoStop}%
\bibitem [{\citenamefont {Wilson}\ \emph {et~al.}(2006)\citenamefont {Wilson},
  \citenamefont {Li}, \citenamefont {Woo}, \citenamefont {Dai}, \citenamefont
  {Mook}, \citenamefont {Frost}, \citenamefont {Komiya},\ and\ \citenamefont
  {Ando}}]{wilson2006high}%
  \BibitemOpen
  \bibfield  {author} {\bibinfo {author} {\bibfnamefont {S.~D.}\ \bibnamefont
  {Wilson}}, \bibinfo {author} {\bibfnamefont {S.}~\bibnamefont {Li}}, \bibinfo
  {author} {\bibfnamefont {H.}~\bibnamefont {Woo}}, \bibinfo {author}
  {\bibfnamefont {P.}~\bibnamefont {Dai}}, \bibinfo {author} {\bibfnamefont
  {H.~A.}\ \bibnamefont {Mook}}, \bibinfo {author} {\bibfnamefont {C.~D.}\
  \bibnamefont {Frost}}, \bibinfo {author} {\bibfnamefont {S.}~\bibnamefont
  {Komiya}}, \ and\ \bibinfo {author} {\bibfnamefont {Y.}~\bibnamefont
  {Ando}},\ }\bibfield  {title} {\enquote {\bibinfo {title} {High-energy spin
  excitations in the electron-doped superconductor
  {P}r$_{0.88}${L}a{C}e$_{0.12}${C}u{O}$_{4\ensuremath{-}\ensuremath{\delta}}$
  with ${T}_{c}=21\text{ }\text{ }\mathrm{K}$},}\ }\href {\doibase
  10.1103/PhysRevLett.96.157001} {\bibfield  {journal} {\bibinfo  {journal}
  {Phys. Rev. Lett.}\ }\textbf {\bibinfo {volume} {96}},\ \bibinfo {pages}
  {157001} (\bibinfo {year} {2006})}\BibitemShut {NoStop}%
\bibitem [{\citenamefont {Coldea}\ \emph {et~al.}(2001)\citenamefont {Coldea},
  \citenamefont {Hayden}, \citenamefont {Aeppli}, \citenamefont {Perring},
  \citenamefont {Frost}, \citenamefont {Mason}, \citenamefont {Cheong},\ and\
  \citenamefont {Fisk}}]{coldea2001spinwaves}%
  \BibitemOpen
  \bibfield  {author} {\bibinfo {author} {\bibfnamefont {R.}~\bibnamefont
  {Coldea}}, \bibinfo {author} {\bibfnamefont {S.~M.}\ \bibnamefont {Hayden}},
  \bibinfo {author} {\bibfnamefont {G.}~\bibnamefont {Aeppli}}, \bibinfo
  {author} {\bibfnamefont {T.~G.}\ \bibnamefont {Perring}}, \bibinfo {author}
  {\bibfnamefont {C.~D.}\ \bibnamefont {Frost}}, \bibinfo {author}
  {\bibfnamefont {T.~E.}\ \bibnamefont {Mason}}, \bibinfo {author}
  {\bibfnamefont {S.-W.}\ \bibnamefont {Cheong}}, \ and\ \bibinfo {author}
  {\bibfnamefont {Z.}~\bibnamefont {Fisk}},\ }\bibfield  {title} {\enquote
  {\bibinfo {title} {Spin waves and electronic interactions in
  {L}a$_2${C}u{O}$_4$},}\ }\href {\doibase 10.1103/PhysRevLett.86.5377}
  {\bibfield  {journal} {\bibinfo  {journal} {Phys. Rev. Lett.}\ }\textbf
  {\bibinfo {volume} {86}},\ \bibinfo {pages} {5377--5380} (\bibinfo {year}
  {2001})}\BibitemShut {NoStop}%
\bibitem [{\citenamefont {Barnes}\ \emph {et~al.}(1993)\citenamefont {Barnes},
  \citenamefont {Dagotto}, \citenamefont {Riera},\ and\ \citenamefont
  {Swanson}}]{barnes1993excitation}%
  \BibitemOpen
  \bibfield  {author} {\bibinfo {author} {\bibfnamefont {T}~\bibnamefont
  {Barnes}}, \bibinfo {author} {\bibfnamefont {E}~\bibnamefont {Dagotto}},
  \bibinfo {author} {\bibfnamefont {J}~\bibnamefont {Riera}}, \ and\ \bibinfo
  {author} {\bibfnamefont {ES}~\bibnamefont {Swanson}},\ }\bibfield  {title}
  {\enquote {\bibinfo {title} {Excitation spectrum of heisenberg spin
  ladders},}\ }\href@noop {} {\bibfield  {journal} {\bibinfo  {journal}
  {Physical Review B}\ }\textbf {\bibinfo {volume} {47}},\ \bibinfo {pages}
  {3196} (\bibinfo {year} {1993})}\BibitemShut {NoStop}%
\bibitem [{\citenamefont {Wang}(1999)}]{wang1999exactly}%
  \BibitemOpen
  \bibfield  {author} {\bibinfo {author} {\bibfnamefont {Yupeng}\ \bibnamefont
  {Wang}},\ }\bibfield  {title} {\enquote {\bibinfo {title} {Exact solution of
  a spin-ladder model},}\ }\href {\doibase 10.1103/PhysRevB.60.9236} {\bibfield
   {journal} {\bibinfo  {journal} {Phys. Rev. B}\ }\textbf {\bibinfo {volume}
  {60}},\ \bibinfo {pages} {9236--9239} (\bibinfo {year} {1999})}\BibitemShut
  {NoStop}%
\bibitem [{\citenamefont {Barnes}\ and\ \citenamefont
  {Riera}(1994)}]{barnes1994susceptibility}%
  \BibitemOpen
  \bibfield  {author} {\bibinfo {author} {\bibfnamefont {Ted}\ \bibnamefont
  {Barnes}}\ and\ \bibinfo {author} {\bibfnamefont {Jos{\'e}}\ \bibnamefont
  {Riera}},\ }\bibfield  {title} {\enquote {\bibinfo {title} {Susceptibility
  and excitation spectrum of ({V}{O})$_2${P}$_2${O}$_7$ in ladder and
  dimer-chain models},}\ }\href@noop {} {\bibfield  {journal} {\bibinfo
  {journal} {Physical Review B}\ }\textbf {\bibinfo {volume} {50}},\ \bibinfo
  {pages} {6817} (\bibinfo {year} {1994})}\BibitemShut {NoStop}%
\bibitem [{\citenamefont {Eschrig}(2006)}]{eschrig2006effect}%
  \BibitemOpen
  \bibfield  {author} {\bibinfo {author} {\bibfnamefont {Matthias}\
  \bibnamefont {Eschrig}},\ }\bibfield  {title} {\enquote {\bibinfo {title}
  {The effect of collective spin-1 excitations on electronic spectra in high-tc
  superconductors},}\ }\href@noop {} {\bibfield  {journal} {\bibinfo  {journal}
  {Advances in Physics}\ }\textbf {\bibinfo {volume} {55}},\ \bibinfo {pages}
  {47--183} (\bibinfo {year} {2006})}\BibitemShut {NoStop}%
\bibitem [{\citenamefont {Wakimoto}\ \emph {et~al.}(2007)\citenamefont
  {Wakimoto}, \citenamefont {Yamada}, \citenamefont {Tranquada}, \citenamefont
  {Frost}, \citenamefont {Birgeneau},\ and\ \citenamefont
  {Zhang}}]{wakimoto2007disappearance}%
  \BibitemOpen
  \bibfield  {author} {\bibinfo {author} {\bibfnamefont {S}~\bibnamefont
  {Wakimoto}}, \bibinfo {author} {\bibfnamefont {K}~\bibnamefont {Yamada}},
  \bibinfo {author} {\bibfnamefont {JM}~\bibnamefont {Tranquada}}, \bibinfo
  {author} {\bibfnamefont {CD}~\bibnamefont {Frost}}, \bibinfo {author}
  {\bibfnamefont {RJ}~\bibnamefont {Birgeneau}}, \ and\ \bibinfo {author}
  {\bibfnamefont {H}~\bibnamefont {Zhang}},\ }\bibfield  {title} {\enquote
  {\bibinfo {title} {Disappearance of antiferromagnetic spin excitations in
  overdoped {L}a$_{2-x}${S}r$_x${C}u{O}$_4$},}\ }\href@noop {} {\bibfield
  {journal} {\bibinfo  {journal} {Phys. Rev. Lett.}\ }\textbf {\bibinfo
  {volume} {98}},\ \bibinfo {pages} {247003} (\bibinfo {year}
  {2007})}\BibitemShut {NoStop}%
\bibitem [{\citenamefont {Kivelson}\ \emph {et~al.}(2003)\citenamefont
  {Kivelson}, \citenamefont {Bindloss}, \citenamefont {Fradkin}, \citenamefont
  {Oganesyan}, \citenamefont {Tranquada}, \citenamefont {Kapitulnik},\ and\
  \citenamefont {Howald}}]{kivelson2003how}%
  \BibitemOpen
  \bibfield  {author} {\bibinfo {author} {\bibfnamefont {S.~A.}\ \bibnamefont
  {Kivelson}}, \bibinfo {author} {\bibfnamefont {I.~P.}\ \bibnamefont
  {Bindloss}}, \bibinfo {author} {\bibfnamefont {E.}~\bibnamefont {Fradkin}},
  \bibinfo {author} {\bibfnamefont {V.}~\bibnamefont {Oganesyan}}, \bibinfo
  {author} {\bibfnamefont {J.~M.}\ \bibnamefont {Tranquada}}, \bibinfo {author}
  {\bibfnamefont {A.}~\bibnamefont {Kapitulnik}}, \ and\ \bibinfo {author}
  {\bibfnamefont {C.}~\bibnamefont {Howald}},\ }\bibfield  {title} {\enquote
  {\bibinfo {title} {How to detect fluctuating stripes in the high-temperature
  superconductors},}\ }\href {\doibase 10.1103/RevModPhys.75.1201} {\bibfield
  {journal} {\bibinfo  {journal} {Rev. Mod. Phys.}\ }\textbf {\bibinfo {volume}
  {75}},\ \bibinfo {pages} {1201--1241} (\bibinfo {year} {2003})}\BibitemShut
  {NoStop}%
\bibitem [{\citenamefont {Fine}(2004)}]{fine2004hypothesis}%
  \BibitemOpen
  \bibfield  {author} {\bibinfo {author} {\bibfnamefont {B.~V.}\ \bibnamefont
  {Fine}},\ }\bibfield  {title} {\enquote {\bibinfo {title} {Hypothesis of
  two-dimensional stripe arrangement and its implications for the
  superconductivity in high-${T}_{c}$ cuprates},}\ }\href {\doibase
  10.1103/PhysRevB.70.224508} {\bibfield  {journal} {\bibinfo  {journal} {Phys.
  Rev. B}\ }\textbf {\bibinfo {volume} {70}},\ \bibinfo {pages} {224508}
  (\bibinfo {year} {2004})}\BibitemShut {NoStop}%
\bibitem [{\citenamefont {Seibold}\ \emph {et~al.}(2011)\citenamefont
  {Seibold}, \citenamefont {Markiewicz},\ and\ \citenamefont
  {Lorenzana}}]{seibold2011spin}%
  \BibitemOpen
  \bibfield  {author} {\bibinfo {author} {\bibfnamefont {G.}~\bibnamefont
  {Seibold}}, \bibinfo {author} {\bibfnamefont {R.~S.}\ \bibnamefont
  {Markiewicz}}, \ and\ \bibinfo {author} {\bibfnamefont {J.}~\bibnamefont
  {Lorenzana}},\ }\bibfield  {title} {\enquote {\bibinfo {title} {Spin canting
  as a result of the competition between stripes and spirals in cuprates},}\
  }\href {\doibase 10.1103/PhysRevB.83.205108} {\bibfield  {journal} {\bibinfo
  {journal} {Phys. Rev. B}\ }\textbf {\bibinfo {volume} {83}},\ \bibinfo
  {pages} {205108} (\bibinfo {year} {2011})}\BibitemShut {NoStop}%
\bibitem [{\citenamefont {Christensen}\ \emph {et~al.}(2007)\citenamefont
  {Christensen}, \citenamefont {R\o{}nnow}, \citenamefont {Mesot},
  \citenamefont {Ewings}, \citenamefont {Momono}, \citenamefont {Oda},
  \citenamefont {Ido}, \citenamefont {Enderle}, \citenamefont {McMorrow},\ and\
  \citenamefont {Boothroyd}}]{christensen2007nature}%
  \BibitemOpen
  \bibfield  {author} {\bibinfo {author} {\bibfnamefont {N.~B.}\ \bibnamefont
  {Christensen}}, \bibinfo {author} {\bibfnamefont {H.~M.}\ \bibnamefont
  {R\o{}nnow}}, \bibinfo {author} {\bibfnamefont {J.}~\bibnamefont {Mesot}},
  \bibinfo {author} {\bibfnamefont {R.~A.}\ \bibnamefont {Ewings}}, \bibinfo
  {author} {\bibfnamefont {N.}~\bibnamefont {Momono}}, \bibinfo {author}
  {\bibfnamefont {M.}~\bibnamefont {Oda}}, \bibinfo {author} {\bibfnamefont
  {M.}~\bibnamefont {Ido}}, \bibinfo {author} {\bibfnamefont {M.}~\bibnamefont
  {Enderle}}, \bibinfo {author} {\bibfnamefont {D.~F.}\ \bibnamefont
  {McMorrow}}, \ and\ \bibinfo {author} {\bibfnamefont {A.~T.}\ \bibnamefont
  {Boothroyd}},\ }\bibfield  {title} {\enquote {\bibinfo {title} {Nature of the
  magnetic order in the charge-ordered cuprate
  {L}a$_{1.48}${N}d$_{0.4}${S}r$_{0.12}${C}u{O}$_{4}$},}\ }\href {\doibase
  10.1103/PhysRevLett.98.197003} {\bibfield  {journal} {\bibinfo  {journal}
  {Phys. Rev. Lett.}\ }\textbf {\bibinfo {volume} {98}},\ \bibinfo {pages}
  {197003} (\bibinfo {year} {2007})}\BibitemShut {NoStop}%
\bibitem [{\citenamefont {Comin}\ \emph {et~al.}(2015)\citenamefont {Comin},
  \citenamefont {Sutarto}, \citenamefont {da~Silva~Neto}, \citenamefont
  {Chauviere}, \citenamefont {Liang}, \citenamefont {Hardy}, \citenamefont
  {Bonn}, \citenamefont {He}, \citenamefont {Sawatzky},\ and\ \citenamefont
  {Damascelli}}]{comin2015broken}%
  \BibitemOpen
  \bibfield  {author} {\bibinfo {author} {\bibfnamefont {R}~\bibnamefont
  {Comin}}, \bibinfo {author} {\bibfnamefont {R}~\bibnamefont {Sutarto}},
  \bibinfo {author} {\bibfnamefont {EH}~\bibnamefont {da~Silva~Neto}}, \bibinfo
  {author} {\bibfnamefont {L}~\bibnamefont {Chauviere}}, \bibinfo {author}
  {\bibfnamefont {R}~\bibnamefont {Liang}}, \bibinfo {author} {\bibfnamefont
  {WN}~\bibnamefont {Hardy}}, \bibinfo {author} {\bibfnamefont
  {DA}~\bibnamefont {Bonn}}, \bibinfo {author} {\bibfnamefont {F}~\bibnamefont
  {He}}, \bibinfo {author} {\bibfnamefont {GA}~\bibnamefont {Sawatzky}}, \ and\
  \bibinfo {author} {\bibfnamefont {A}~\bibnamefont {Damascelli}},\ }\bibfield
  {title} {\enquote {\bibinfo {title} {Broken translational and rotational
  symmetry via charge stripe order in underdoped
  {Y}{B}a$_{2}${C}u$_{3}$o$_{6+y}$},}\ }\href@noop {} {\bibfield  {journal}
  {\bibinfo  {journal} {Science}\ }\textbf {\bibinfo {volume} {347}},\ \bibinfo
  {pages} {1335--1339} (\bibinfo {year} {2015})}\BibitemShut {NoStop}%
\bibitem [{\citenamefont {Fine}(2016)}]{fine2015comment}%
  \BibitemOpen
  \bibfield  {author} {\bibinfo {author} {\bibfnamefont {B}~\bibnamefont
  {Fine}},\ }\bibfield  {title} {\enquote {\bibinfo {title} {Comment on
  ``{B}roken translational and rotational symmetry via charge stripe order in
  underdoped {Y}{B}a$_{2}${C}u$_{3}$o$_{6+y}$''},}\ }\href@noop {} {\bibfield
  {journal} {\bibinfo  {journal} {Science}\ }\textbf {\bibinfo {volume}
  {351}},\ \bibinfo {pages} {235} (\bibinfo {year} {2016})}\BibitemShut
  {NoStop}%
\bibitem [{\citenamefont {Comin}\ \emph {et~al.}(2016)\citenamefont {Comin},
  \citenamefont {Sutarto}, \citenamefont {da~Silva~Neto}, \citenamefont
  {Chauviere}, \citenamefont {Liang}, \citenamefont {Hardy}, \citenamefont
  {Bonn}, \citenamefont {He}, \citenamefont {Sawatzky},\ and\ \citenamefont
  {Damascelli}}]{comin2016reply}%
  \BibitemOpen
  \bibfield  {author} {\bibinfo {author} {\bibfnamefont {R}~\bibnamefont
  {Comin}}, \bibinfo {author} {\bibfnamefont {R}~\bibnamefont {Sutarto}},
  \bibinfo {author} {\bibfnamefont {EH}~\bibnamefont {da~Silva~Neto}}, \bibinfo
  {author} {\bibfnamefont {L}~\bibnamefont {Chauviere}}, \bibinfo {author}
  {\bibfnamefont {R}~\bibnamefont {Liang}}, \bibinfo {author} {\bibfnamefont
  {WN}~\bibnamefont {Hardy}}, \bibinfo {author} {\bibfnamefont
  {DA}~\bibnamefont {Bonn}}, \bibinfo {author} {\bibfnamefont {F}~\bibnamefont
  {He}}, \bibinfo {author} {\bibfnamefont {GA}~\bibnamefont {Sawatzky}}, \ and\
  \bibinfo {author} {\bibfnamefont {A}~\bibnamefont {Damascelli}},\ }\bibfield
  {title} {\enquote {\bibinfo {title} {Reply to comment on ``broken
  translational and rotational symmetry via charge stripe order in underdoped
  {Y}{B}a$_{2}${C}u$_{3}$o$_{6+y}$''},}\ }\href@noop {} {\bibfield  {journal}
  {\bibinfo  {journal} {Science}\ }\textbf {\bibinfo {volume} {351}},\ \bibinfo
  {pages} {235} (\bibinfo {year} {2016})}\BibitemShut {NoStop}%
\bibitem [{\citenamefont {Manousakis}(1991)}]{manousakis1991spin}%
  \BibitemOpen
  \bibfield  {author} {\bibinfo {author} {\bibfnamefont {Efstratios}\
  \bibnamefont {Manousakis}},\ }\bibfield  {title} {\enquote {\bibinfo {title}
  {The spin-$1/2$ heisenberg antiferromagnet on a square lattice and its
  application to the cuprous oxides},}\ }\href@noop {} {\bibfield  {journal}
  {\bibinfo  {journal} {Reviews of Modern Physics}\ }\textbf {\bibinfo {volume}
  {63}},\ \bibinfo {pages} {1} (\bibinfo {year} {1991})}\BibitemShut {NoStop}%
\bibitem [{\citenamefont {Misumi}\ \emph {et~al.}(2014)\citenamefont {Misumi},
  \citenamefont {Seki},\ and\ \citenamefont {Ohta}}]{misumi2014spin}%
  \BibitemOpen
  \bibfield  {author} {\bibinfo {author} {\bibfnamefont {K}~\bibnamefont
  {Misumi}}, \bibinfo {author} {\bibfnamefont {K}~\bibnamefont {Seki}}, \ and\
  \bibinfo {author} {\bibfnamefont {Y}~\bibnamefont {Ohta}},\ }\bibfield
  {title} {\enquote {\bibinfo {title} {Spin excitations in the square-lattice
  heisenberg model with ring-exchange interactions},}\ }in\ \href@noop {}
  {\emph {\bibinfo {booktitle} {Proceedings of the International Conference on
  Strongly Correlated Electron Systems (SCES2013)}}}\ (\bibinfo {year} {2014})\
  p.\ \bibinfo {pages} {014021}\BibitemShut {NoStop}%
\bibitem [{\citenamefont {Lorenzana}\ \emph {et~al.}(2005)\citenamefont
  {Lorenzana}, \citenamefont {Seibold},\ and\ \citenamefont
  {Coldea}}]{lorenzana2005sum}%
  \BibitemOpen
  \bibfield  {author} {\bibinfo {author} {\bibfnamefont {J}~\bibnamefont
  {Lorenzana}}, \bibinfo {author} {\bibfnamefont {G}~\bibnamefont {Seibold}}, \
  and\ \bibinfo {author} {\bibfnamefont {R}~\bibnamefont {Coldea}},\ }\bibfield
   {title} {\enquote {\bibinfo {title} {Sum rules and missing spectral weight
  in magnetic neutron scattering in the cuprates},}\ }\href@noop {} {\bibfield
  {journal} {\bibinfo  {journal} {Physical Review B}\ }\textbf {\bibinfo
  {volume} {72}},\ \bibinfo {pages} {224511} (\bibinfo {year}
  {2005})}\BibitemShut {NoStop}%
\bibitem [{\citenamefont {Lieb}\ and\ \citenamefont
  {Mattis}(1962)}]{lieb1962ordering}%
  \BibitemOpen
  \bibfield  {author} {\bibinfo {author} {\bibfnamefont {Elliott}\ \bibnamefont
  {Lieb}}\ and\ \bibinfo {author} {\bibfnamefont {Daniel}\ \bibnamefont
  {Mattis}},\ }\bibfield  {title} {\enquote {\bibinfo {title} {Ordering energy
  levels of interacting spin systems},}\ }\href@noop {} {\bibfield  {journal}
  {\bibinfo  {journal} {Journal of Mathematical Physics}\ }\textbf {\bibinfo
  {volume} {3}},\ \bibinfo {pages} {749--751} (\bibinfo {year}
  {1962})}\BibitemShut {NoStop}%
\bibitem [{\citenamefont {Hakobyan}(2008)}]{hakobyan2008antiferromagnetic}%
  \BibitemOpen
  \bibfield  {author} {\bibinfo {author} {\bibfnamefont {Tigran}\ \bibnamefont
  {Hakobyan}},\ }\bibfield  {title} {\enquote {\bibinfo {title}
  {Antiferromagnetic ordering of energy levels for a spin ladder with four-spin
  cyclic exchange: Generalization of the lieb-mattis theorem},}\ }\href@noop {}
  {\bibfield  {journal} {\bibinfo  {journal} {Physical Review B}\ }\textbf
  {\bibinfo {volume} {78}},\ \bibinfo {pages} {012407} (\bibinfo {year}
  {2008})}\BibitemShut {NoStop}%
\bibitem [{\citenamefont {Lieb}(1989)}]{lieb1989two}%
  \BibitemOpen
  \bibfield  {author} {\bibinfo {author} {\bibfnamefont {Elliott~H}\
  \bibnamefont {Lieb}},\ }\bibfield  {title} {\enquote {\bibinfo {title} {Two
  theorems on the hubbard model},}\ }\href@noop {} {\bibfield  {journal}
  {\bibinfo  {journal} {Physical Review Letters}\ }\textbf {\bibinfo {volume}
  {62}},\ \bibinfo {pages} {1201} (\bibinfo {year} {1989})}\BibitemShut
  {NoStop}%
\bibitem [{\citenamefont {Tranquada}\ \emph {et~al.}(1989)\citenamefont
  {Tranquada}, \citenamefont {Shirane}, \citenamefont {Keimer}, \citenamefont
  {Shamoto},\ and\ \citenamefont {Sato}}]{tranquada1989neutron}%
  \BibitemOpen
  \bibfield  {author} {\bibinfo {author} {\bibfnamefont {J.~M.}\ \bibnamefont
  {Tranquada}}, \bibinfo {author} {\bibfnamefont {G.}~\bibnamefont {Shirane}},
  \bibinfo {author} {\bibfnamefont {B.}~\bibnamefont {Keimer}}, \bibinfo
  {author} {\bibfnamefont {S.}~\bibnamefont {Shamoto}}, \ and\ \bibinfo
  {author} {\bibfnamefont {M.}~\bibnamefont {Sato}},\ }\bibfield  {title}
  {\enquote {\bibinfo {title} {Neutron scattering study of magnetic excitations
  in {Y}{B}a$_{2}${C}u$_{3}$o$_{6+x}$},}\ }\href {\doibase
  10.1103/PhysRevB.40.4503} {\bibfield  {journal} {\bibinfo  {journal} {Phys.
  Rev. B}\ }\textbf {\bibinfo {volume} {40}},\ \bibinfo {pages} {4503--4516}
  (\bibinfo {year} {1989})}\BibitemShut {NoStop}%
\bibitem [{\citenamefont {Hayden}\ \emph {et~al.}(1996)\citenamefont {Hayden},
  \citenamefont {Aeppli}, \citenamefont {Perring}, \citenamefont {Mook},\ and\
  \citenamefont {Do\ifmmode~\breve{g}\else \u{g}\fi{}an}}]{hayden1996high}%
  \BibitemOpen
  \bibfield  {author} {\bibinfo {author} {\bibfnamefont {S.~M.}\ \bibnamefont
  {Hayden}}, \bibinfo {author} {\bibfnamefont {G.}~\bibnamefont {Aeppli}},
  \bibinfo {author} {\bibfnamefont {T.~G.}\ \bibnamefont {Perring}}, \bibinfo
  {author} {\bibfnamefont {H.~A.}\ \bibnamefont {Mook}}, \ and\ \bibinfo
  {author} {\bibfnamefont {F.}~\bibnamefont {Do\ifmmode~\breve{g}\else
  \u{g}\fi{}an}},\ }\bibfield  {title} {\enquote {\bibinfo {title}
  {High-frequency spin waves in {Y}{B}a$_{2}${C}u$_{3}$o$_{6.15}$},}\ }\href
  {\doibase 10.1103/PhysRevB.54.R6905} {\bibfield  {journal} {\bibinfo
  {journal} {Phys. Rev. B}\ }\textbf {\bibinfo {volume} {54}},\ \bibinfo
  {pages} {R6905--R6908} (\bibinfo {year} {1996})}\BibitemShut {NoStop}%
\bibitem [{\citenamefont {T.~Egami}\ and\ \citenamefont
  {Singh}(2010)}]{egami2010spin-lattice}%
  \BibitemOpen
  \bibfield  {author} {\bibinfo {author} {\bibfnamefont {D.~Parshall
  A.~Subedi}\ \bibnamefont {T.~Egami}, \bibfnamefont {B.~V.~Fine}}\ and\
  \bibinfo {author} {\bibfnamefont {D.~J.}\ \bibnamefont {Singh}},\ }\bibfield
  {title} {\enquote {\bibinfo {title} {Spin-lattice coupling and
  superconductivity in {F}e pnictides},}\ }\href@noop {} {\bibfield  {journal}
  {\bibinfo  {journal} {Adv. Cond. Mat. Phys.}\ }\textbf {\bibinfo {volume}
  {2010}},\ \bibinfo {pages} {164916} (\bibinfo {year} {2010})}\BibitemShut
  {NoStop}%
\bibitem [{\citenamefont {Egami}\ \emph {et~al.}(2010)\citenamefont {Egami},
  \citenamefont {Fine}, \citenamefont {Singh}, \citenamefont {Parshall},
  \citenamefont {de~La~Cruz},\ and\ \citenamefont {Dai}}]{egami2010spin}%
  \BibitemOpen
  \bibfield  {author} {\bibinfo {author} {\bibfnamefont {T.}~\bibnamefont
  {Egami}}, \bibinfo {author} {\bibfnamefont {B.~V.}\ \bibnamefont {Fine}},
  \bibinfo {author} {\bibfnamefont {D.~J.}\ \bibnamefont {Singh}}, \bibinfo
  {author} {\bibfnamefont {D.}~\bibnamefont {Parshall}}, \bibinfo {author}
  {\bibfnamefont {C.}~\bibnamefont {de~La~Cruz}}, \ and\ \bibinfo {author}
  {\bibfnamefont {P.}~\bibnamefont {Dai}},\ }\bibfield  {title} {\enquote
  {\bibinfo {title} {Spin--lattice coupling in iron-pnictide
  superconductors},}\ }\href {https://doi.org/10.1016/j.physc.2009.11.167}
  {\bibfield  {journal} {\bibinfo  {journal} {Physica C: Superconductivity and
  its applications}\ }\textbf {\bibinfo {volume} {470}},\ \bibinfo {pages}
  {S294--S295} (\bibinfo {year} {2010})}\BibitemShut {NoStop}%
\bibitem [{\citenamefont {Hayden}\ \emph {et~al.}(2004)\citenamefont {Hayden},
  \citenamefont {Mook}, \citenamefont {Dai}, \citenamefont {Perring},\ and\
  \citenamefont {Do{\u{g}}an}}]{hayden2004structure}%
  \BibitemOpen
  \bibfield  {author} {\bibinfo {author} {\bibfnamefont {S.~M.}\ \bibnamefont
  {Hayden}}, \bibinfo {author} {\bibfnamefont {H.~A.}\ \bibnamefont {Mook}},
  \bibinfo {author} {\bibfnamefont {P.}~\bibnamefont {Dai}}, \bibinfo {author}
  {\bibfnamefont {T.~G.}\ \bibnamefont {Perring}}, \ and\ \bibinfo {author}
  {\bibfnamefont {F.}~\bibnamefont {Do{\u{g}}an}},\ }\bibfield  {title}
  {\enquote {\bibinfo {title} {The structure of the high-energy spin
  excitations in a high-transition-temperature superconductor},}\ }\href@noop
  {} {\bibfield  {journal} {\bibinfo  {journal} {Nature}\ }\textbf {\bibinfo
  {volume} {429}},\ \bibinfo {pages} {531--534} (\bibinfo {year}
  {2004})}\BibitemShut {NoStop}%
\bibitem [{\citenamefont {Stock}\ \emph {et~al.}(2005)\citenamefont {Stock},
  \citenamefont {Buyers}, \citenamefont {Cowley}, \citenamefont {Clegg},
  \citenamefont {Coldea}, \citenamefont {Frost}, \citenamefont {Liang},
  \citenamefont {Peets}, \citenamefont {Bonn}, \citenamefont {Hardy},\ and\
  \citenamefont {Birgeneau}}]{stock2005from}%
  \BibitemOpen
  \bibfield  {author} {\bibinfo {author} {\bibfnamefont {C.}~\bibnamefont
  {Stock}}, \bibinfo {author} {\bibfnamefont {W.~J.~L.}\ \bibnamefont
  {Buyers}}, \bibinfo {author} {\bibfnamefont {R.~A.}\ \bibnamefont {Cowley}},
  \bibinfo {author} {\bibfnamefont {P.~S.}\ \bibnamefont {Clegg}}, \bibinfo
  {author} {\bibfnamefont {R.}~\bibnamefont {Coldea}}, \bibinfo {author}
  {\bibfnamefont {C.~D.}\ \bibnamefont {Frost}}, \bibinfo {author}
  {\bibfnamefont {R.}~\bibnamefont {Liang}}, \bibinfo {author} {\bibfnamefont
  {D.}~\bibnamefont {Peets}}, \bibinfo {author} {\bibfnamefont
  {D.}~\bibnamefont {Bonn}}, \bibinfo {author} {\bibfnamefont {W.~N.}\
  \bibnamefont {Hardy}}, \ and\ \bibinfo {author} {\bibfnamefont {R.~J.}\
  \bibnamefont {Birgeneau}},\ }\bibfield  {title} {\enquote {\bibinfo {title}
  {From incommensurate to dispersive spin-fluctuations: The high-energy
  inelastic spectrum in superconducting {Y}{B}a$_{2}${C}u$_{3}${O}$_{6.5}$},}\
  }\href {\doibase 10.1103/PhysRevB.71.024522} {\bibfield  {journal} {\bibinfo
  {journal} {Phys. Rev. B}\ }\textbf {\bibinfo {volume} {71}},\ \bibinfo
  {pages} {024522} (\bibinfo {year} {2005})}\BibitemShut {NoStop}%
\end{thebibliography}%

\pagebreak

\onecolumngrid

\begin{center}
\textbf{\Large Supplemental material}\\
\medskip

\end{center}
\setcounter{equation}{0}
\setcounter{figure}{0}
\setcounter{table}{0}
\setcounter{page}{1}
\makeatletter
\renewcommand{\theequation}{S\arabic{equation}}
\renewcommand{\thefigure}{S\arabic{figure}}

\renewcommand{\theHtable}{Supplement.\thetable}
\renewcommand{\theHfigure}{Supplement.\thefigure}



Here, for the convenience of readers, we reproduce several experimental plots of spin response in cuprates obtained in neutron scattering experiments.

\begin{figure*}[h]
\includegraphics[width = \textwidth]
{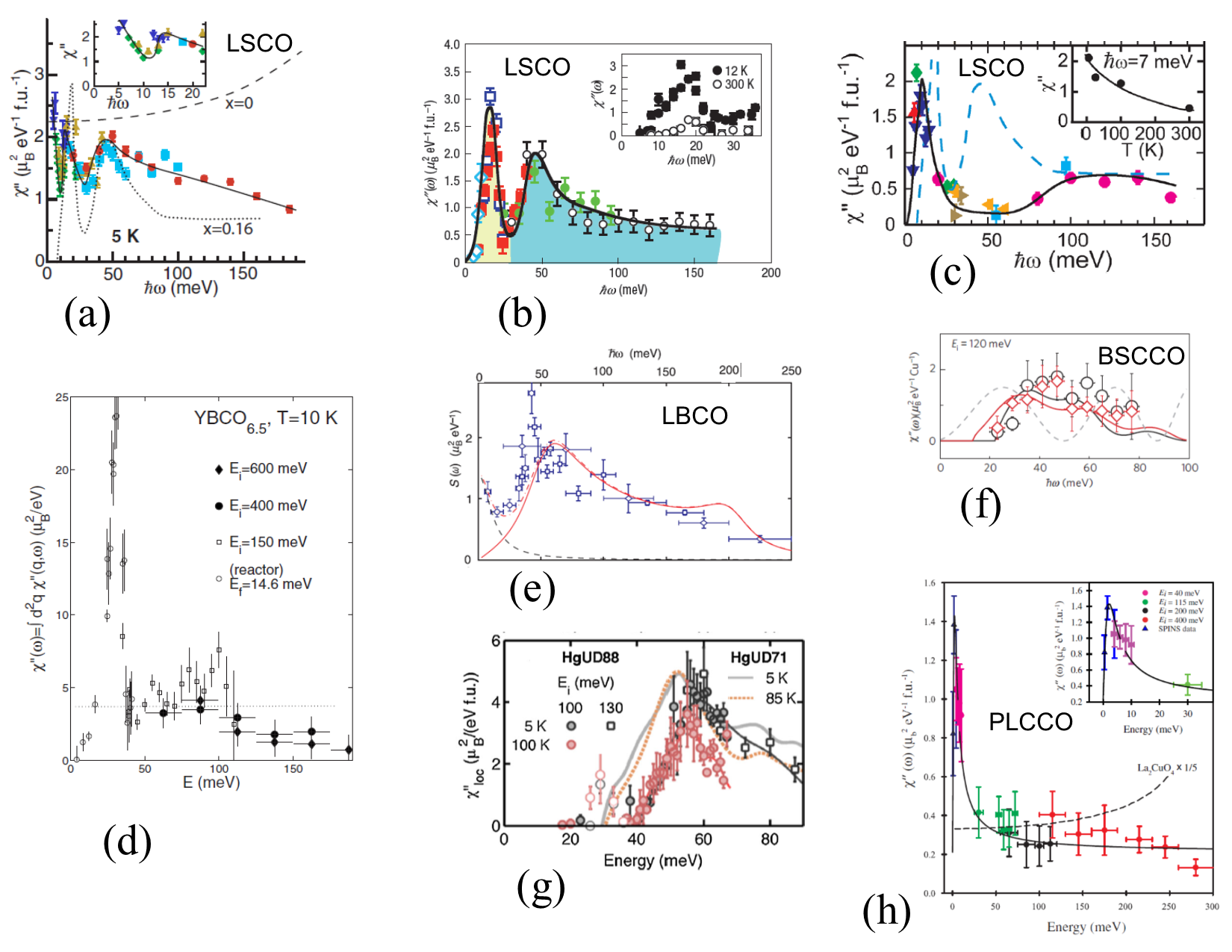}
%
 \caption{\label{fig:experimental}
 Experimental plots for the $\Q$-integrated magnetic susceptibility $\chi''(\omega)$ in various cuprates.
(a)-(c):  underdoped\cite{lipscombe2009emergence} ($x=0.085$), optimally dopped\cite{vignolle2007two} ($x=0.16$) and overdoped\cite{lipscombe2007persistence} ($x=0.22$) La$_{2-x}$Sr$_x$CuO$_4$;
%
(d) YBa$_2$Cu$_3$O$_{6.5}$, \cite{stock2010effect}
(e) La$_{1.875}$Ba$_{0.125}$CuO$_4$, \cite{tranquada2004quantum}
(f) optimally dopped Bi$_2$Sr$_2$CaCu$_2$O$_{8+\delta}$,  \cite{xu2009testing}
(g) underdoped and optimally doped  HgBa$_2$CuO$_{4+\delta}$ \cite{chan2016hourglass,chan2016commensurate} and
(h) electron-doped ${\mathrm{Pr}}_{0.88}{\mathrm{LaCe}}_{0.12}{\mathrm{CuO}}_{4\ensuremath{-}\ensuremath{\delta}}.$\cite{wilson2006high} For the detailed description of the plot legends see the original references.  Adapted from the indicated references with permissions from the authors and copyright holders. 
}
\end{figure*}

\newpage

\begin{figure*}[t]
\includegraphics[width = 0.35 \textwidth]{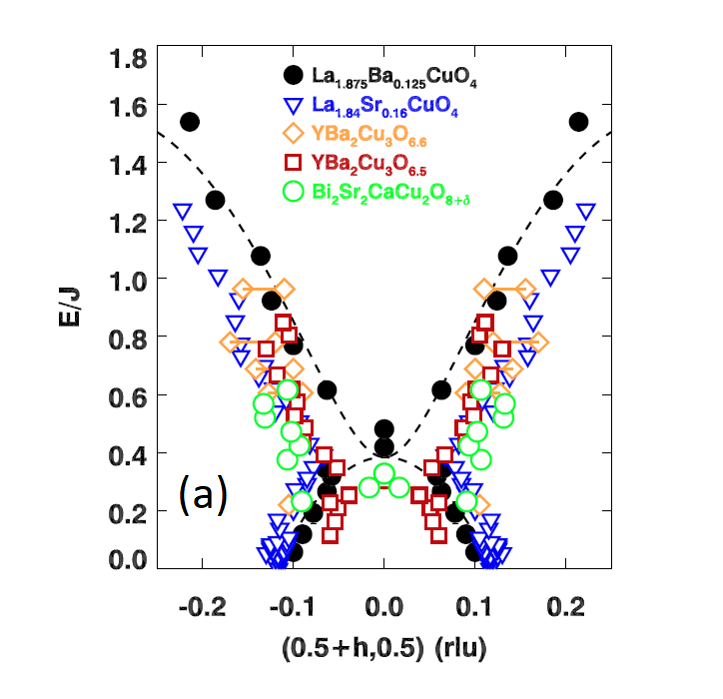}
\includegraphics[width = 0.2\textwidth]{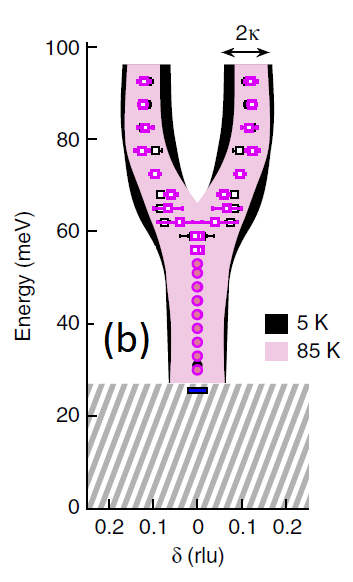}
\includegraphics[width = 0.23\textwidth]{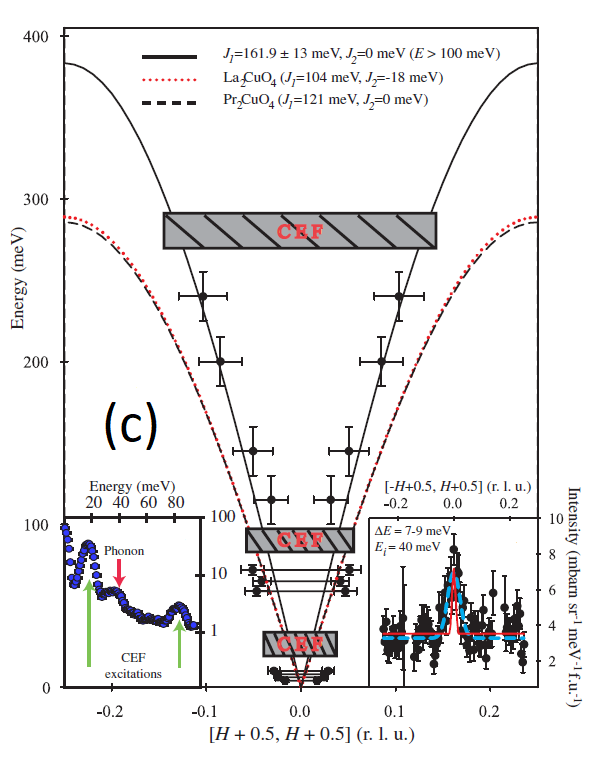}
 \caption{(a) Hourglass spectrum in various cuprates, figure from Ref. \onlinecite{fujita2011progress}. Shown is the section of $\chi''\left(\Q,\omega\right)$ at $\Q=(\pi+2\pi h,2\pi h)$ for La$_{1.875}$Ba$_{0.125}$CuO$_4$,\cite{tranquada2004quantum} La$_{1.84}$Sr$_{0.16}$CuO$_4$,\cite{vignolle2007two}
 YBa$_2$Cu$_3$O$_{6.6}$,\cite{hayden2004structure}  YBa$_2$Cu$_3$O$_{6.5}$,\cite{stock2005from,stock2010effect} and  Bi$_2$Sr$_2$CaCu$_2$O$_{8+\delta}$\cite{xu2009testing} (b) Y-shaped spectrum in  underdoped HgBa$_2$CuO$_{4+x}$\cite{chan2016commensurate}. 
 (c) Y-shaped spectrum in electron-doped ${\mathrm{Pr}}_{0.88}{\mathrm{LaCe}}_{0.12}{\mathrm{CuO}}_{4\ensuremath{-}\ensuremath{\delta}}.$\cite{wilson2006high} For the detailed description of the plot legends see the original references.  Adapted from the indicated references with permissions from the authors and copyright holders. 
\label{fig:hourglass}
 }
\end{figure*}

\noindent {\it Acknowledgements.}
We are grateful to the authors and copyright holders of refs. 1 -  10 for permitting us to reproduce their figures.

\end{document}